\documentclass[twocolumn,aps,prb,eqsecnum,showpacs,floatfix]{revtex4}
\usepackage{graphicx}
\usepackage{dcolumn}
\usepackage{bm}
\usepackage{amsfonts}
\usepackage{amssymb}
\usepackage{amsmath}
\graphicspath{{./figures/}}

\newcommand{\ocite}{\onlinecite}

\newcommand{\iy}{\infty}

\newcommand{\dg}{\dagger}

\newcommand{\lt}{\left}
\newcommand{\rt}{\right}

\newcommand{\fr}{\frac}

\newcommand{\sq}{\sqrt}
\newcommand{\lbl}{\label}

\newcommand{\ot}{\otimes}

\newcommand{\eq}[1]{Eq.~(\ref{eq:#1})}
\newcommand{\eqs}[2]{Eqs.~(\ref{eq:#1}) and (\ref{eq:#2})}

\newcommand{\secr}[1]{Sec.~\ref{sec:#1}}

\newcommand{\figr}[1]{Fig.~\ref{fig:#1}}

\newcommand{\beq}{\begin{equation}}
\newcommand{\eeq}{\end{equation}}
\newcommand{\bfg}{\begin{figure}}
\newcommand{\efg}{\end{figure}}

\newcommand{\bsp}[1]{\begin{split}#1\end{split}}
\newcommand{\bal}[1]{\begin{align}#1\end{align}}
\newcommand{\beqar}{\begin{eqnarray*}}
\newcommand{\eeqar}{\end{eqnarray*}}
\newcommand{\beqarn}{\begin{eqnarray}}
\newcommand{\eeqarn}{\end{eqnarray}}
\newcommand{\ba}{\begin{array}}
\newcommand{\ea}{\end{array}}
\newcommand{\bwt}{\begin{widetext}}
\newcommand{\ewt}{\end{widetext}}

\newcommand{\sgn}{{\rm sgn}\,}

\newcommand{\ra}{\rightarrow}

\newcommand{\Ra}{\Rightarrow}

\newcommand{\lra}{\leftrightarrow}

\newcommand{\Ht}{\tilde{H}}

\newcommand{\kb}{{\bf k}}

\newcommand{\pb}{{\bf p}}

\newcommand{\ga}{\gamma}
\newcommand{\Ga}{\Gamma}

\newcommand{\sig}{\sigma}
\begin{document}
\title{Exotic surface states in hybrid structures of topological insulators and Weyl semimetals}
\author{Stefan~Juergens and Bj\"orn Trauzettel}
\affiliation{Institute of Theoretical Physics and Astrophysics, University of W\"urzburg, D-97074 W\"urzburg, Germany}
\date{\today}
\begin{abstract}
Topological insulators (TIs) and Weyl semimetals (WSMs) are two realizations of topological matter usually appearing 
separately in nature. However, they are directly related to each other via a topological phase transition. In this paper, we investigate the question whether these two topological phases can exist together at the same time, with a combined, hybrid surface state at the joint boundaries. We analyze effective models of a 3D TI and an inversion symmetric WSM 
and couple them in a way that certain symmetries, like inversion, are preserved. A tunnel coupling approach enables us to obtain the hybrid surface state Hamiltonian analytically. This offers the possibility of a detailed study of its dispersion relation depending on the investigated couplings. For 
spin-symmetric coupling, we find that two Dirac nodes can emerge out of the combination of a single Dirac node and a Fermi arc. For spin-asymmetric coupling, the dispersion relation is gapped and the former Dirac node gets spin-polarized. 
We propose different experimental realization of the hybrid system, including compressively strained HgTe as well as heterostructures of TI and WSM materials. 
\end{abstract}

\maketitle

\section{Introduction}\lbl{sec:intro}

The study of topological properties in a semiconductor environment has become a strong and flourishing field in condensed matter physics. 
Topological insulators (TIs) are the standard materials in this context, well studied both theoretically 
and experimentally by now~\cite{Hasan2010,Qi2011,Ando2013}. Their semi-metallic counter-parts, Weyl semimetals (WSMs)~\cite{Hosur2013,Turner2013,Jia2016,Yan2017}, were also proposed to exist 
in condensed matter systems decades ago~\cite{Nielsen1981,Nielsen1983,Murakami2007}. However, only very recently with the prediction of 
concrete material realizations~\cite{Wan2011,Weng2015,Huang2015} the field has seen an enormous growth. The experimental proof of the existence of Weyl points 
and their corresponding surface states, called Fermi arcs, followed soon afterwards~\cite{Xu2015,Lv2015,Lv2015b,Yang2015,Xu2015b,Lu2015}. 
Yet both for fundamental research and application purposes these "early" WSMs, such as the TaAs family of non-centrosymmetric monopnictides, are 
too complicated with many Weyl points (24 for TaAs) in the Brillouin zone. Simpler materials with eight~\cite{Chang2015,Ruan2016,Rauch2015,Ruan2016b,Sun2015,Huang2016,Deng2016,Tamai2016,Jiang2016,Xu2016} 
and four~\cite{Koepernik2016,Belopolski2016} Weyl points have been predicted and observed, where the 
latter is the minimal number of Weyl points for a system with time-reversal symmetry (TRS). Materials with broken TRS~\cite{Borisenko2015} could realize the absolute 
minimum of two Weyl points, but for that case only theoretical proposals\cite{Burkov2011,Cho2011,Xu2011,Bulmash2014,Wang2016} exist so far. 
Most of them rely on magnetically doped TIs or TI heterostructures. 

TI and WSM are both topological phases that can be directly connected to each other through quantum phase transitions~\cite{Murakami2007,Okugawa2014}. 
In this paper we want to go a step further and study the question whether a system can be both in the TI and WSM phase at the same time, or at least 
support both corresponding surface states, 2D Dirac surface states and Fermi arcs, on the same surface. 

Such a combined phase might exist in HgTe with applied compressive strain. The strain pushes the $\Ga_8$ bands into 
one another, creating Dirac points which are then split by breaking of inversion symmetry through bulk inversion asymmetry (BIA) terms~\cite{Ruan2016}. 
At the same time, the topological band inversion between the $\Ga_8$ and the $\Ga_6$ bands remains, leading 
to the conjecture that this system could have topological Dirac states and Fermi arcs on its surface. 

A different way to create such a hybrid surface state is placing a TI and WSM spatially adjacent to each other, 
possibly separated by a small, topological trivial buffer layer. The separate surface states of TI and WSM 
will interact, e.g. by Coulomb interaction or tunneling due to a small overlap of wave functions, forming the hybrid surface dispersion relation. 
Previous related research on adjacent TI and WSM phases~\cite{Grushin2015} suggest that at such a shared 
surface both Dirac states and Fermi arcs can exist. However, they were found in different areas of $k$-space, mutually excluding one another 
such that they do not hybridize at all. Our approach differs from the one chosen in Ref.~\onlinecite{Grushin2015} by considering only a small, perturbative coupling between the 
two phases. This ensures that both TI and WSM surface states survive and can interact with each other. 

We focus in this paper on an analytical study of the combined surface states generated from the hybridized TI and WSM. A simplified ansatz 
offers the possibility to calculate the surface Hamiltonian analytically, allowing for a detailed analysis of the surface physics. Depending 
on the symmetry of the assumed couplings, the surface dispersion relation shows quite different behavior. In the case of spin symmetry, 
two shifted Dirac nodes may emerge out of the combination of a single Dirac node and a Fermi arc. For spin-asymmetric coupling, 
the Fermi arc gaps out and spin-polarizes the former Dirac node. 

The article is organized as follows. We recap effective models 
for the separate phases of TIs and WSMs~\cite{Zhang2009,Liu2010,Yang2011,Okugawa2014,McCormick2016} 
and discuss their symmetry properties and surface states in \secr{model}. 
The coupling of the Hamiltonians and the analytic form of the surface state is discussed in \secr{coupled_sys}. 
\secr{surface_disp} focuses on the different ways to influence and tune the combined surface dispersion relation. 
Possible experimental realizations are proposed in \secr{experiment}. 

\section{Separate Models}\lbl{sec:model}

The model of the TI phase we will use was originally derived for the Bi$_2$Se$_3$ family of materials 
in Refs.~\ocite{Zhang2009, Liu2010}. It contains four bands and serves as a minimal, but general, TI model. 
The Weyl Hamiltonian considered in the following originates from Refs.~\ocite{Yang2011,McCormick2016}. It contains 
two bands and models an inversion symmetric type I or II WSM with two Weyl points. 
We simplify the models as far as possible without 
losing too much versatility. It is important to retain terms quadratic in momentum for the introduction of the 
surface in the $z$ direction. This is done via hardwall boundary conditions on a half space $z\leq0$ or $z\geq0$. 

\subsection{Topological Insulator}\lbl{sec:model_TI}

The effective Hamiltonian for a 3D TI is given by the 4x4 matrix~\cite{Zhang2009,Liu2010}
\beq
    H_{TI} = \lt(\ba{cc}M(k)\tau_3+Bk_z\tau_2+C\tau_0 & iAk_{-}\tau_1 \\ -iA^*k_{+}\tau_1 & M(k)\tau_3+Bk_z\tau_2+C\tau_0 \ea\rt)
\lbl{eq:H_TI}
\eeq
with $M(k)=M_0+M_1\lt(k_\Vert^2+k_z^2\rt)$, $k_\Vert^2=k_x^2+k_y^2$ and $k_{\pm}=k_x\pm ik_y=k_\Vert e^{\pm i\phi_k}$. 
In the original derivation for Bi$_2$Se$_3$ the Pauli matrices $\vec{\tau}$ describe an orbital degree of freedom. 
$H_{TI}$ is written in a spin-up/down basis, represented by the Pauli matrices $\vec{\sig}$ in the following. The coupling 
$A=\lt|A\rt|e^{i\phi_A}$ can in principle be complex. 
The model is in the strong TI phase for $M_0M_1<0$.

We define the inversion operator $P_{TI}=\sig_0\ot\tau_3$ and 
time-reversal operator $T_{TI}=i\sig_2\ot\tau_0K$ with $K$ the 
complex conjugation operator. $H_{TI}$ is symmetric under both operations, fulfilling 
\beq \bsp{
	P_{TI}^{\dg}H_{TI}\lt(-k\rt)P_{TI} &= H_{TI}\lt(k\rt), \\
	\ T_{TI}^{\dg}H_{TI}\lt(-k\rt)T_{TI} &= H_{TI}\lt(k\rt).\lbl{eq:sym_TI}
} \eeq
The bulk dispersion relation is double degenerate and given by
\beq
	E_{TI}=C\pm\sq{\lt|A\rt|^{2}k_\Vert^{2}+B^{2}k_{z}^{2}+M(k)^{2}}. \lbl{eq:E_TI}
\eeq
Based on the method described in App.~\ref{sec:app_boundary}, 
the surface states can be calculated analytically. We assume opposite surfaces to be well separated, which offers the possibility to 
treat them individually. Thus in the calculation we only consider one of them via hardwall boundary conditions at $z=0$. 
The surface wave function is then given by
\beq
	\Psi\lt(z\rt)=\fr{1}{\sq{2}}\lt(e^{ik_{z,1}z}-e^{ik_{z,2}z}\rt)\lt(\ba{c} 
	\pm\fr{ i\eta Ak_{-}}{\lt|A\rt|k_\Vert} \psi_\eta \\ \psi_\eta \ea\rt) \lbl{eq:simple_ansatz_TI}
\eeq
with $\psi_\eta=\fr{1}{\sq{2}}\lt(\ba{c} 1 \\ \eta \ea\rt)$ and the inverse localization length 
$ik_{z,\substack{1\\2}}=\fr{1}{2M_1} \lt[-\eta B \pm\sq{4M_1\lt(M_0+M_1 k_\Vert^2\rt)+B^2}\rt]$. 
The sign $\eta=\pm$ depends on the surface, 
$\eta=- \sgn\lt(\fr{B}{M_1}\rt)$ (upper surface) or $\eta=\sgn\lt(\fr{B}{M_1}\rt)$ (lower surface). 

The existence condition for the surface state, see App.~\ref{sec:app_boundary}, is 
\beq
	M_1\lt(M_0+M_1 k_\Vert^2\rt)<0 \lbl{eq:exist_cond_TI}
\eeq
stressing the importance of being in the inverted regime. 

The surface Hamiltonian (dispersion relation) is obtained from $H_{TI}$ by projecting out the orbital (orbital \& spin) 
degrees of freedom with the help of $\psi_\eta$ ($\Psi\lt(z\rt)$). We find the usual Dirac form 
\beq
	H_{TI}^{sur}=\lt(\ba{cc} C & i\eta Ak_{-} \\ -i\eta A^*k_{+} & C \ea\rt);\ E_{TI}^{sur}=C \pm \lt|A\rt|k_\Vert,\lbl{eq:E_TI_sur}
\eeq
experiencing spin-momentum locking, with the angle $\phi_A+\fr{\eta\pi}{2}$ between the spin projection and momentum vector in the $x$-$y$ plane. 
The combined dispersion relations of the bulk and surface of the TI are shown in \figr{TI_disp}.
\bfg
	\includegraphics[width=0.4\textwidth]{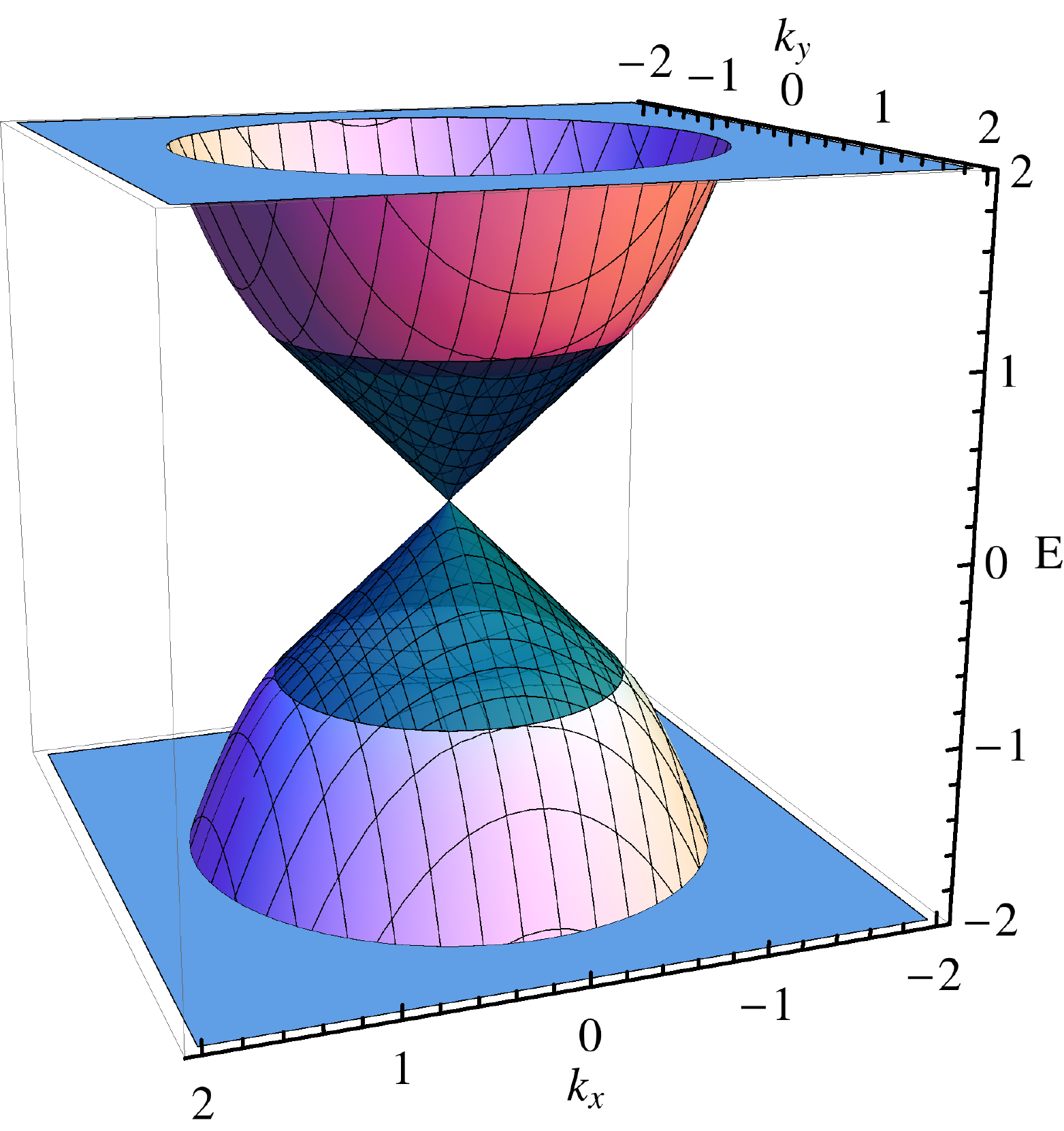}
	\caption{Bulk and surface dispersion relations, \eqs{E_TI}{E_TI_sur}, of the TI model. The surface band is plotted in cyan. \\
	Parameters: $C=0$, $M_0=-1$, $M_1=1$, $A=1$, $B=1$, $k_z=0$.}
	\lbl{fig:TI_disp}
\efg

\subsection{Inversion symmetric Weyl Semimetal}\lbl{sec:model_W}

A WSM exists in different flavors. On the one hand, one distinguishes type I and type II depending 
on preserved or broken Lorentz invariance at the Weyl points~\cite{Xu2015c,Soluyanov2015,Sharma2016}. Secondly, 
either time-reversal or inversion symmetry has to be broken to get from a Dirac to a Weyl semimetal. For all these phases 
minimal models have been proposed in the literature~\cite{Yang2011,Okugawa2014,Dwivedi2016,McCormick2016}. 

For simplicity, we focus on the model with broken time-reversal and preserved inversion symmetry, as it has 
the minimal number of one pair of Weyl points. The Hamiltonian is
\beq
    H_{W} = t(k)\tau_3+v_zk_z\tau_2+v_yk_y\tau_1+\ga t\lt(k_x^2-k_W^2\rt)\tau_0
\lbl{eq:H_W}
\eeq
with $t(k)=t\lt(k_\Vert^2+k_z^2-k_W^2\rt)$. The degree of freedom described by the 
Pauli matrices $\vec{\tau}$ can be orbital, spin or a combination of the two, depending 
on the specific material realization. For the concrete form of the symmetry operations considered in the following 
we assume a spinless system~\cite{McCormick2016}. The two Weyl points are specified by $k_x=\pm k_W$. The parameter $\ga$ leads to a tilting of 
the dispersion relation at the Weyl points. For $|\ga|<1$ one has a type I, otherwise a type II WSM. 
Expanding $H_{W}$ around $k_{x}=\pm k_{W}$ yields a Hamiltonian with linearized Weyl form
\beq
	H_{W}^{lin}=v_yk_{y}\tau_{1}+v_zk_{z}\tau_{2}\pm 2tk_Wk_{x}\lt(\tau_{3}+\ga\tau_{0}\rt).
\eeq
The Hamiltonian $H_{W}$ fulfills the symmetry conditions 
\beq \bsp{
	P_W^{\dg}H_{W}\lt(-k\rt)P_W &= H_{W}\lt(k\rt), \\
	T_W^{\dg}H_{W}\lt(-k\rt)T_W &\neq H_{W}\lt(k\rt)\lbl{eq:sym_W}
} \eeq
 with the inversion operator $P_W=\tau_{3}$ and time-reversal operator
$T_W=\tau_{0}K$ with $K$ the complex conjugation operator. Hence, parity is preserved and time-reversal symmetry broken. 

The bulk dispersion relation is then given by
\beq
	E_{W}=\ga t\lt(k_x^2-k_W^2\rt)\pm\sq{v_y^{2}k_{y}^{2}+v_z^2k_{z}^{2}+t(k)^{2}}. \lbl{eq:E_W}
\eeq
The surface states can be calculated analytically based on the method discussed 
in the App.~\ref{sec:app_boundary}. Their wave function is given by
\beq
	\Psi\lt(z\rt)=\lt(e^{ik_{z,1}z}-e^{ik_{z,2}z}\rt) \psi_\eta \lbl{eq:simple_ansatz_W}
\eeq
with $\psi_\eta=\fr{1}{\sq{2}}\lt(\ba{c} 1 \\ \eta \ea\rt)$ and inverse localization length 
$ik_{z,\substack{1\\2}}=\fr{1}{2t} \lt[-\eta v_z \pm\sq{4t^2 \lt(k_\Vert^2-k_W^2\rt)+v_z^2}\rt]$. The sign $\eta=\pm$ depends on the surface; 
$\eta=- \sgn\lt(\fr{v_z}{t}\rt)$ (upper surface) or $\eta=\sgn\lt(\fr{v_z}{t}\rt)$ (lower surface). 

The existence condition for the surface state is 
\beq
	k_\Vert^2 < k_W^2 \lbl{eq:exist_cond_W}
\eeq
such that Fermi arcs can only exist between the Weyl points. 

Hence, the surface dispersion relation yields the known Fermi arc spectrum
\beq
	E_{W}^{sur}=\ga t\lt(k_x^2-k_W^2\rt) + \eta v_y k_y.\lbl{eq:E_W_sur}
\eeq
The combined dispersion relations of the bulk and surface of the WSM are shown in \figr{W_disp}.
\bfg
	\includegraphics[width=0.4\textwidth]{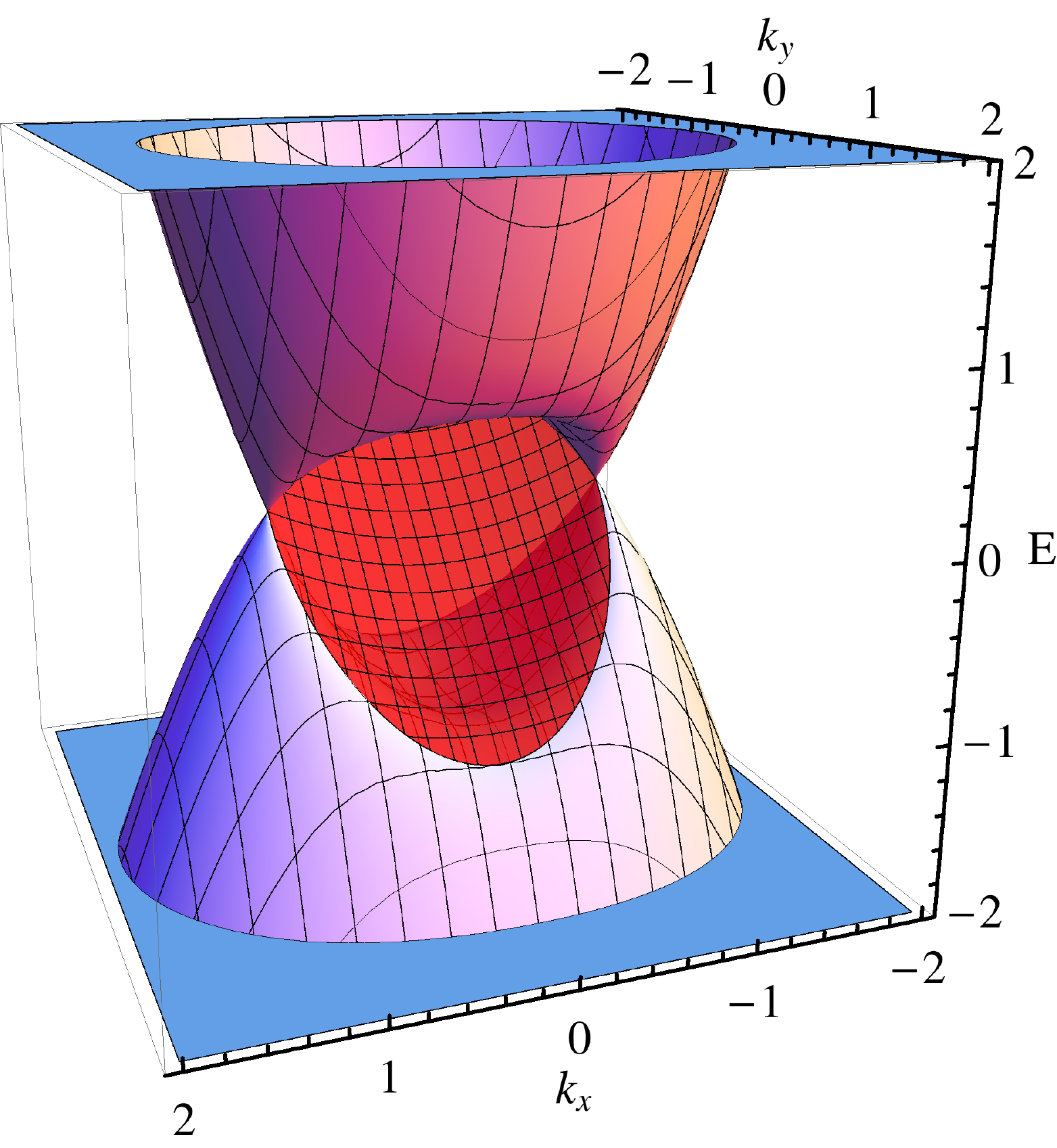}
	\caption{Bulk and upper surface dispersion relations, \eqs{E_W}{E_W_sur}, of the Weyl model. The surface band is plotted in red. \\ 
	Parameters: $\ga=\fr{1}{4}$, $k_W=1$, $t=1$, $v_y=1$, $v_z=1$, $k_z=0$.}
	\lbl{fig:W_disp}
\efg

\section{Coupled system}\lbl{sec:coupled_sys}

The Hamiltonians and surface wave functions of the TI and WSM phases 
discussed in \secr{model} are very similar. Thus, we 
conjecture that also the combined system may have surface states which can be calculated by the 
simplified method described in App.~\ref{sec:app_boundary}. This will 
allow us to discuss the surface physics analytically. 

In this section, we define the combined Hamiltonian and discuss the couplings 
allowed by symmetry under the assumptions that certain symmetries are preserved. 
The surface state Hamiltonian and wave function are derived 
and the limitations due to the approximated calculation method are discussed. 

The combined Hamiltonian of the TI and WSM phases is defined by
\beq
	H_{WTI}=\lt(\ba{cc} H_{TI} & H_{C} \\ H_{C}^{\dg} & H_{W} \ea\rt)\lbl{eq:H_WTI}
\eeq
with the coupling $H_C$. Such a coupling can be regarded as a tunneling Hamiltonian approach 
where $H_C$ (weakly) couples the two entities $H_{TI}$ and $H_{W}$. A similar approach has been 
considered in Ref.~\onlinecite{Baum2015} to combine topological systems of different kinds with each other 
and study their emerging physics. The combined symmetry operator for inversion symmetry is now given by
\beq
	P_{WTI}=\lt(\ba{cc} P_{TI} & 0 \\ 0 & P_{W} \ea\rt). 
\eeq
As time-reversal symmetry is already broken in the subsystem of the WSM, 
it will also be absent in the combined system. The study of the effect on the TI of such a breaking of 
time-reversal symmetry via coupling, applicable e.g. in the setup of spatially separate Weyl and TI phases as depicted 
in \figr{exp_realization} (b), is one of the goals of this paper. The stability of gapless edge states to time-reversal symmetry breaking perturbations such as magnetic fields\cite{Ma2015} 
and considerable Coulomb interaction\cite{Kharitonov2016} is an active research topic and has been experimentally observed in 2D. It is proposed that 
crystalline symmetries such as inversion or rotational symmetries protect the gapless edge states in the absence of time-reversal symmetry. As inversion symmetry is preserved 
in our system, we conjecture that the use of the gapless TI model can be justified even in a time-reversal breaking environment.  

Applying the inversion operator to the 
Hamiltonian, following \eqs{sym_TI}{sym_W}, yields restrictions for the allowed 
couplings assuming that this symmetry is not broken. As the symmetry operator is block-diagonal, these restrictions do not depend 
on $H_W$ or $H_{TI}$. 

For an inversion symmetric system, the couplings proportional to $\tau_3$ and $\tau_0$ have to be even in momentum, 
while the ones proportional to $\tau_2$ and $\tau_1$ have to be odd in momentum. We choose the following representation 
\beq
	H_{C,IS}=\lt(\ba{c} H_{c,IS} \\ \Ht_{c,IS} \ea\rt)
\eeq
with
\beq
	H_{c,IS}=d(k_\Vert)\tau_3+c_1k_+\tau_2 + b_1k_+\tau_1+a(k_\Vert)\tau_0,\lbl{eq:H_c}
\eeq
where $d(k_\Vert)=d_0+d_2k_\Vert^2$, $a(k_\Vert)=a_0+a_2k_\Vert^2$,
and $\Ht_{c,IS}$ having the same structure. This choice ensures the preservation of parity for the combined system. 
The size of the terms depends on the concrete experimental realization, where the best candidate materials 
for our proposal have yet to be identified. In the case of two spatially separate Weyl and TI systems, as 
depicted in \figr{exp_realization} (b), the coupling parameters can be calculated from the overlap of the wave functions of 
the different materials. As an example, this is done in Ref.~\onlinecite{Michetti2012} 
for a bilayer HgTe quantum well system by fitting a $\kb\cdot\pb$ model to experimentally obtained band 
structures. In general, all symmetry allowed couplings can be relevant for the following discussion. 

In this paper however, coupling terms proportional to $k_z$ are not considered, for simplicity. 
This is a physically reasonable assumption at least for the surface states if one assumes them to be 2D, 
perfectly localized in the $z$-direction. Close to the Weyl points or the TI bulk band edge where the surface states delocalize a $k_z$ dependent 
coupling should be taken into account. 

The ansatz we will consider is 
\beq
	\Psi\lt(z\rt)=\lt(e^{ik_{z,1}z}-e^{ik_{z,2}z}\rt)\lt(\ba{c} L_1\lt(k_{\pm}\rt) \psi_{\eta_{TI}} \\ 
	L_2\lt(k_{\pm}\rt) \psi_{\eta_{TI}} \\ L_3\lt(k_{\pm}\rt) \psi_{\eta_W} \ea\rt) \lbl{eq:simple_ansatz_WTI}
\eeq
with $\psi_\eta=\fr{1}{\sq{2}}\lt(\ba{c} 1 \\ \eta \ea\rt)$. This is a special case of the general form of the surface wave function 
$\Psi_g\lt(z\rt)=\sum_ja_je^{ik_{z,j}z}\psi\lt(k_{\pm},k_{z,j}\rt)$, $j\in\{1,...,6\}$. Its choice 
is motivated by the ability to obtain analytical solutions for the surface states. 
Physically it means that we only consider solutions were the TI and WSM surface states have the 
same exponential localization with the same localization length. This implies that phase transitions of 
the subsystems, such as normal insulator (NI) to TI or NI to WSM, can not be discussed separatly in this treatment. 
However, for a system deep in the TI and WSM phase, the simplification should not alter the essential physics. We have 
checked numerically that small differences in the localization lengths of the subsystems do not alter the surface dispersion relations in a qualitative way, 
see App. \ref{sec:app_numerics}. 

Projecting the Hamiltonian \eq{H_WTI} on the surface, this separates 
the eigenvalue equation into simpler problems 
\bal{
	H_{WTI}^{sur}\lt(\ba{c} L1 \\ L2 \\ L3 \ea \rt) &= E_{WTI}^{sur}\lt(\ba{c} L1 \\ L2 \\ L3 \ea\rt), 
	\lbl{eq:eig_eq_WTI_disp} \\
	H_{WTI}^{k_z}\lt( \ba{c} L1 \\ L2 \\ L3 \ea\rt) &= 0
	\lbl{eq:eig_eq_WTI_kz}
}
with the Hamiltonians
\bwt
\beqarn
	H_{WTI}^{sur} &=& \lt(\ba{ccc} C & i\eta Ak_{-} & a(k_\Vert)+\eta b_{1}k_{+} \\
	-i\eta A^*k_{+} & C & \tilde{a}(k_\Vert)+\eta \tilde{b}_{1}k_{+} \\
	a(k_\Vert)^*+\eta b_{1}^*k_{-} & \tilde{a}(k_\Vert)^*+\eta 
	\tilde{b}_{1}^*k_{-} & \ga t\lt(k_x^2-k_W^2\rt)+\eta v_yk_{y} \ea\rt),
	\lbl{eq:H_WTI^sur} \\
	& \underset{part.\ diag.}{\Ra}& \lt(\ba{ccc} C + \lt|A\rt|k_\Vert & 0 & \Ht_{c}+e^{i\phi^A_{k}}H_{c}\\ 
	0 & C - \lt|A\rt|k_\Vert & \Ht_{c}-e^{i\phi^A_{k}}H_{c}\\ 
	\Ht^*_{c}+e^{-i\phi^A_{k}}H^*_{c}& 
	\Ht^*_{c}-e^{-i\phi^A_{k}}H^*_{c}& \ga t\lt(k_x^2-k_W^2\rt)+\eta v_yk_{y} \ea\rt),
	\lbl{eq:H_WTI^sur_2} \\
	H_{WTI}^{k_z} &=& \lt(\ba{ccc} M\lt(k\rt)-i\eta Bk_{z} & 0 & d(k_\Vert)-i\eta c_{1}k_{+} \\
 	0 & M\lt(k\rt)-i\eta Bk_{z} & \tilde{d}(k_\Vert)-i\eta \tilde{c}_{1}k_{+} \\
	d(k_\Vert)^*-i\eta c_{1}^*k_{-} & \tilde{d}(k_\Vert)^*-i\eta \tilde{c}_{1}^*k_{-} & t\lt(k\rt)-i\eta v_z k_{z} \ea\rt) \lbl{eq:H_WTI^k_z}
\eeqarn
\ewt
for $\eta=\eta_{TI}=\eta_{W}$. In \eq{H_WTI^sur_2}, we partially diagonalize the Hamiltonian and define 
$\phi^A_{k}=\phi_k-\phi_A-\eta\fr{\pi}{2}$, $H_{c}=a(k_\Vert)+\eta b_{1}k_{+}$ and $\Ht_{c}=\tilde{a}(k_\Vert)+\eta \tilde{b}_{1}k_{+}$. 
This will help in the interpretation of the surface dispersion relation in terms of coupled Dirac cone and Fermi arc. 
In the case of $\eta=\eta_{TI}=-\eta_{W}$, one 
has to replace in Eqs.~(\ref{eq:H_WTI^sur}) - (\ref{eq:H_WTI^k_z}) 
$a(k_\Vert)\lra d(k_\Vert)$, $b_1\lra i c_1$, $v_y\ra -v_y$ and $v_z\ra -v_z$. We will 
focus in the following on the former, $\eta_{TI}=\eta_{W}$, case. 
Taking $\lt(\ba{ccc} L1 & L2 & L3 \ea\rt)^T$ as the same eigenvector in \eqs{eig_eq_WTI_disp}{eig_eq_WTI_kz}, 
the latter can only be fulfilled by further restrictions on the parameters. 
We choose a locking between some of the TI and the WSM parameters, i.e. $t(k)=\nu M(k)$ and $v_z=\nu B$ with 
$\nu$ a constant (set to 1 in the following). This ensures the same localization length 
for the two subsystems. Additionally, the couplings $c_1$ and $d(k_\Vert)$ are set to be 0 for simplicity. 
Therefore the total coupling does not change the original orbital character of the TI and WSM 
surface states, being eigenstates of the $\tau_1$ matrix with fixed eigenvalue $+$ or $-$. 
With regard to these restrictions, we have checked that the neglected couplings can be considered numerically with only quantitative changes 
to the surface dispersion relations, see App. \ref{sec:app_numerics}. 

In total, this leads to the same 
quadratic equation for $k_z$ as in the pure TI case, $ik_{z,\substack{1\\2}}=\fr{1}{2M_1} \lt[-\eta B \pm\sq{4M_1\lt(M_0+M_1 k_\Vert^2\rt)+B^2}\rt]$. 
The existence condition is again
\beq
	M_1\lt(M_0+M_1 k_\Vert^2\rt)<0 \lbl{eq:exist_cond_WTI}
\eeq
and the (unnormalized) eigenvectors are given by
\beq
	\lt( \ba{c} L1 \\ L2 \\ L3 \ea\rt)=\lt( \ba{c} 
	\lt(E_{WTI}^{sur}-C\rt)H_{c}+i\eta Ak_-\Ht_c \\ 
	\lt(E_{WTI}^{sur}-C\rt)\Ht_c-i\eta A^*k_+H_{c} \\ 
	\lt(E_{WTI}^{sur}-C\rt)^2-\lt|A\rt|^2k_\Vert^2 \ea\rt).
\eeq
The eigenenergies $E_{WTI}^{sur}$ are too lengthy to state them here, but can also be derived analytically. 

The obtained solution leads to the possibility to tune bulk and surface dispersion relations rather independently. Parameters $M_i$ and 
$B$ influence the surface dispersion relation only indirectly via the existence condition and finite $\ga$ parameter, while 
they strongly influence the bulk band structure as will be shown in the next section. Tuning the coupling 
constants $v_y$, $A$, $a\lt(k_\Vert\rt)$, $b_1$ and their relative phase will still provide a rich parameter space to be explored below. 

\section{Surface dispersion relation}\lbl{sec:surface_disp}

In this section, we discuss the influence of the different coupling parameters on the combined surface states of TIs and WSMs. 
Depending on the choice of symmetries of the coupling, observed phenomenas are the generation of additional Dirac points in the dispersion relation 
or the spin polarization of certain surface bands. 

\subsection{Uncoupled scenario}\lbl{sec:uncoup_scen}

Beginning with the uncoupled case, $H_{C,IS}=0$, the dispersion relations of the surface and bulk states are shown in \figr{surf_disp_uncoup}. 
\bfg%[!]
	\includegraphics[width=0.45\textwidth]{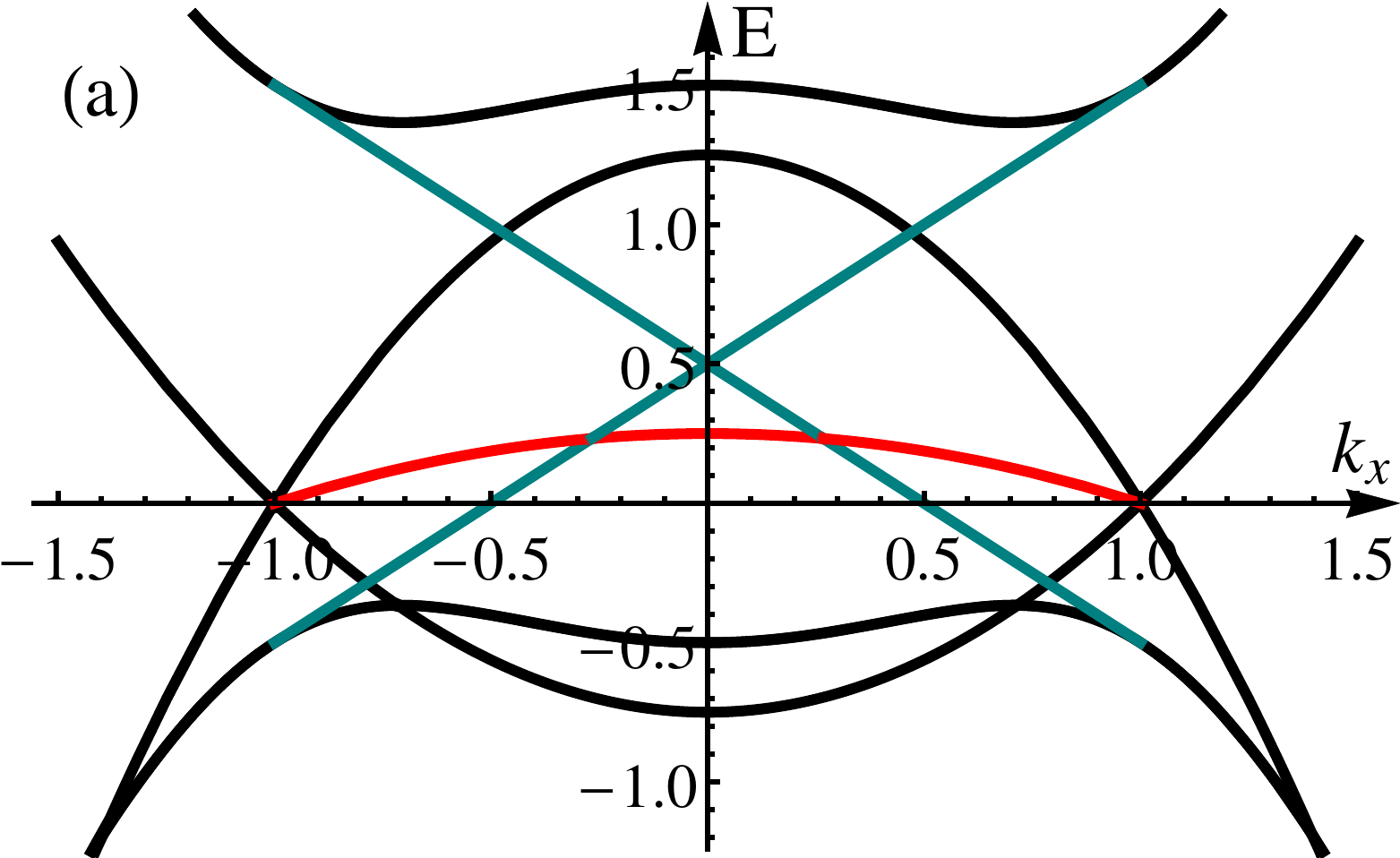}
	\includegraphics[width=0.45\textwidth]{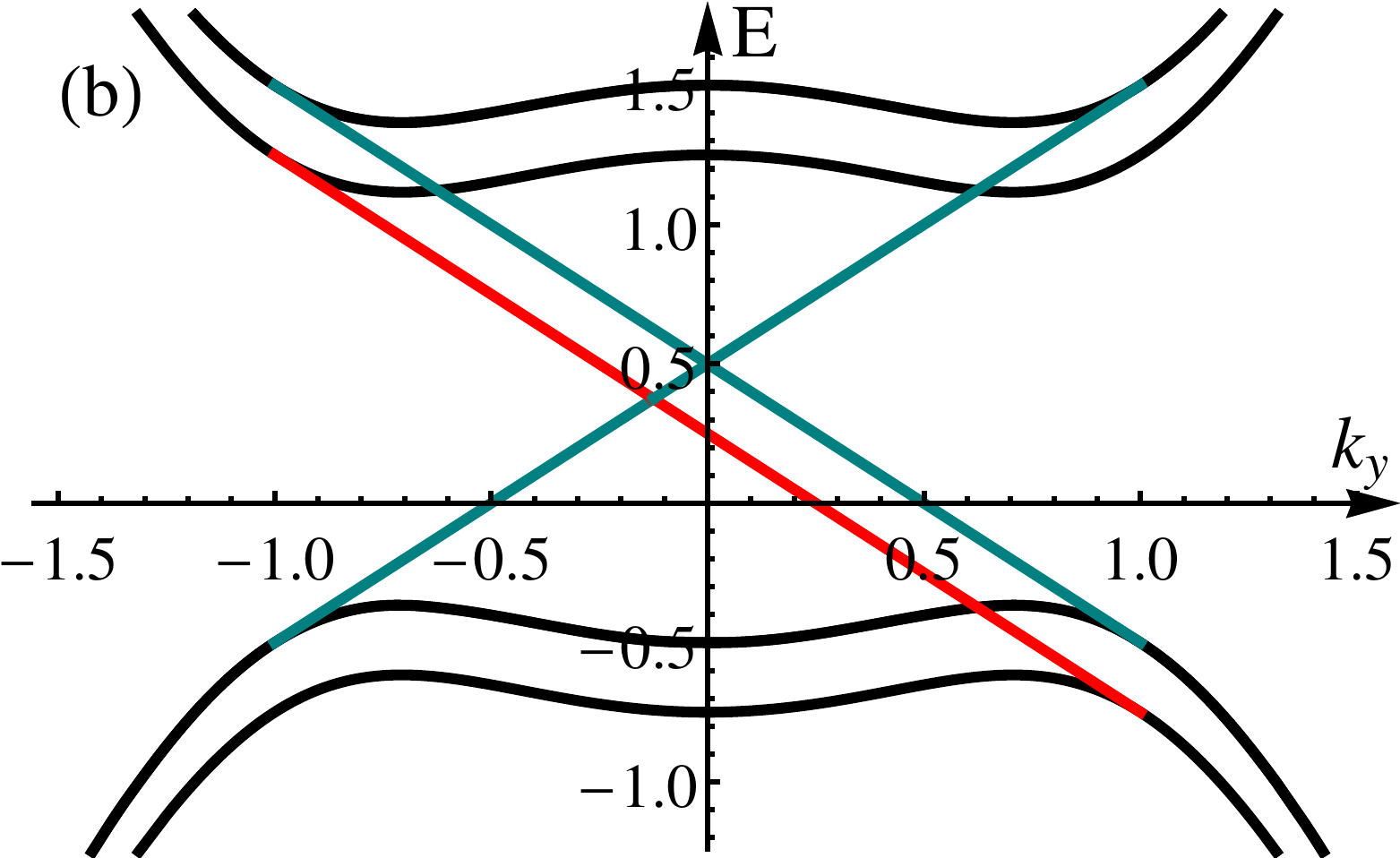}
	\includegraphics[width=0.4\textwidth]{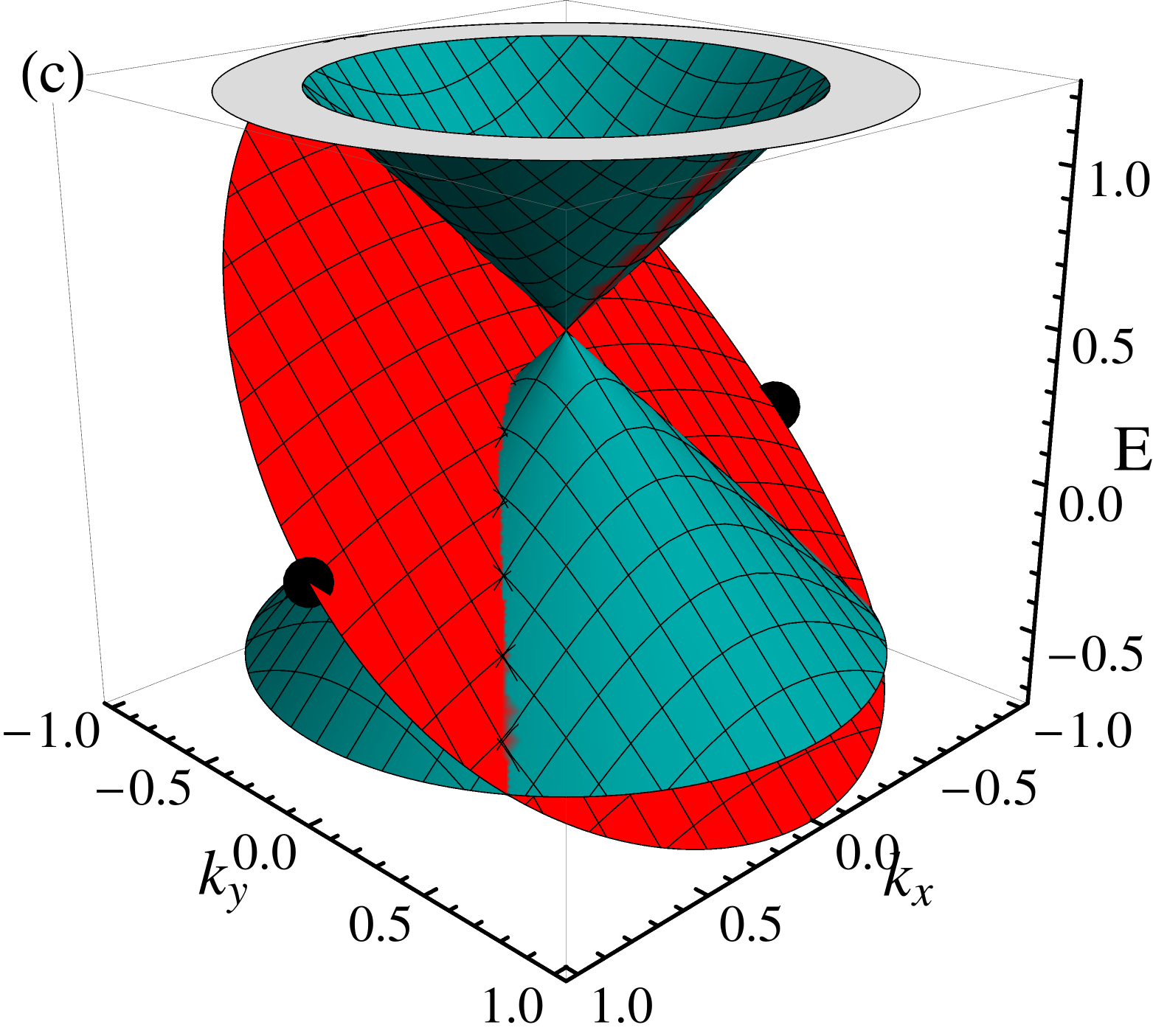}
	\caption{(a) and (b) Bulk and upper surface dispersion relations of the uncoupled TI and WSM models. 
	Color code: Black lines for the bulk states, red for WSM character and cyan for the TI character of the surface states.
	(c) 3D plot of the surface dispersion relation. The two black dots denote the position of the bulk Weyl points.  \\ 
	Parameters: $C=\fr{1}{2}$, $M_0=-1$, $M_1=1$, $B=1$, $k_z=0$, $\ga=-\fr{1}{4}$, $A=1$, $v_y=1$, $a(k_\Vert)=b_1=0$, $H_{c}=\Ht_{c}$.}
	\lbl{fig:surf_disp_uncoup}
\efg
The black lines denote the bulk dispersion relation, cyan (from blue (green) for spin up (down)) and red stand for the TI and WSM surface states, respectively. The 
two black dots give the position of the bulk Weyl points. 
We note that the surface states originate at the bulk states, but cross them unaffectedly. Together with the fact that 
one can tune the bulk gap $M_0$ without changing the surface dispersion relation (aside from the existence condition), we find the possibility to discuss the bulk and surface dispersion relations 
rather separately from each other. It will always be possible to increase the bulk gap and the distance between the two Weyl points
such that the interesting surface physics happens in regions of the Brillouin zone where no bulk state is located. Therefore, 
we will focus in the following on tuning of the surface dispersion relation only. 
In numerical calculations, see App. \ref{sec:app_numerics}, purely exponentially decaying 
surface states do not coexist with bulk states at the same energy and momenta. This is due to finite hybridization between the bulk and surface states. 

\subsection{Real, spin-symmetric coupling: Creation of additional Dirac points}\lbl{sec:real_sym_coup}

A straight-forward way to couple TI and WSM is a real and spin-symmetric coupling via $a(k_\Vert)>0$ or $b_1>0$ 
with $H_{c}=\Ht_{c}$. This kind of coupling leads generally to two Dirac points in the combined surface dispersion relation, as 
plotted in \figr{surf_disp_a0}. 
\bfg
	\includegraphics[width=0.45\textwidth]{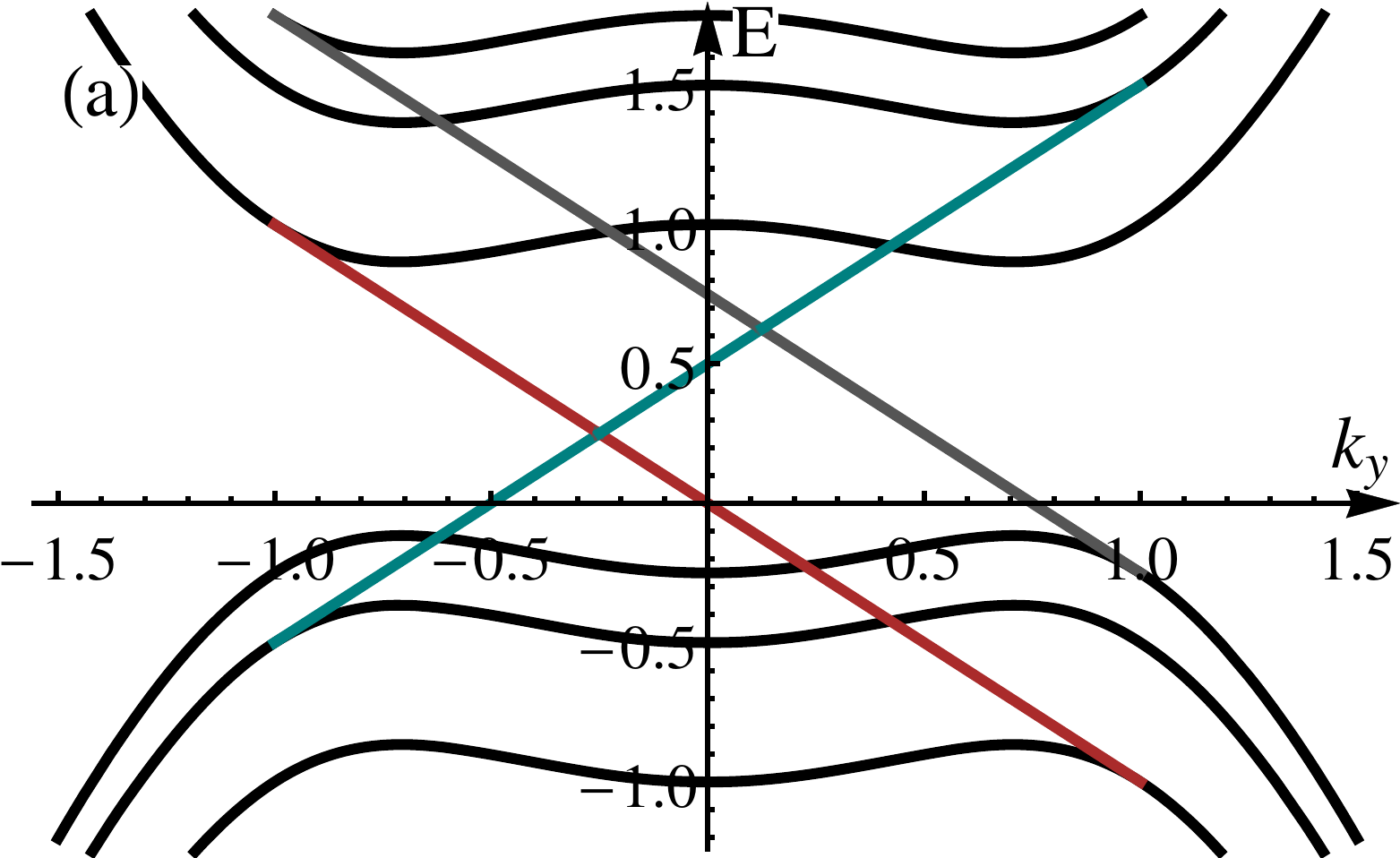}
	\includegraphics[width=0.4\textwidth]{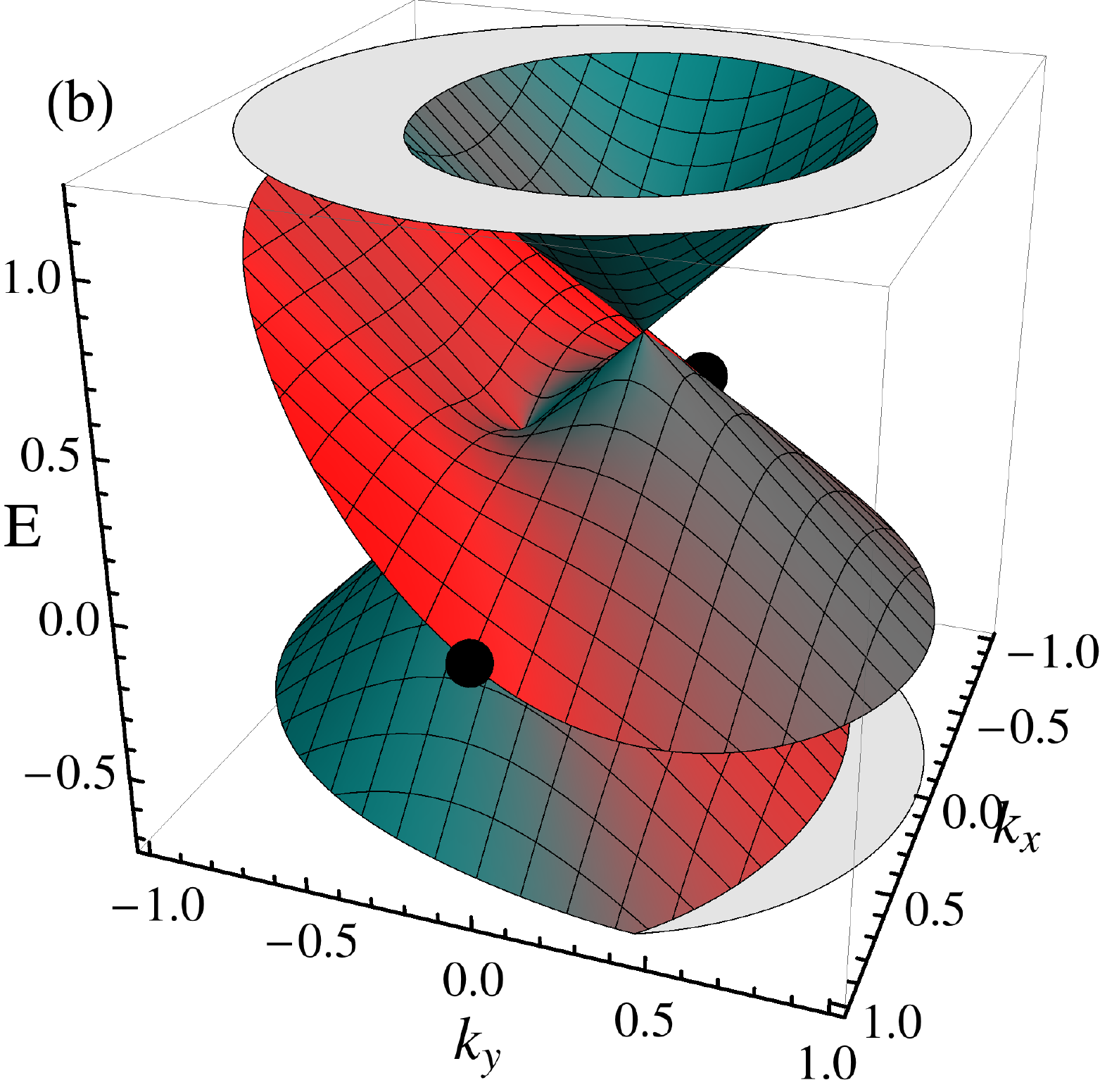}
	\caption{(a) Bulk and upper surface dispersions relation of the TI \& WSM model with real, spin-symmetric coupling. 
	Color code: Black lines for the bulk states, red for WSM character and cyan for the TI character of the surface states.
	(b) 3D plot of the surface dispersion relation. Two Dirac points are visible. \\ 	
	Parameters: $C=\fr{1}{2}$, $M_0=-1$, $M_1=1$, $B=1$, $k_z=0$, $\ga=-\fr{1}{4}$, $A=1$, $v_y=1$, $a(k_\Vert)=\fr{1}{4}$, $b_1=0$, $H_{c}=\Ht_{c}$.}
	\lbl{fig:surf_disp_a0}
\efg
One Dirac point is just shifted by the coupling to the Weyl surface state. The other one is created out of the Weyl and Dirac states 
along a momentum direction where there is no coupling between these two bands. 
Under the assumption that both spin species 
couple equally strong to the WSM, $|\Ht_{c}|=|H_{c}|$, there is always such a momentum direction $\phi_k$ 
where one part (hole or electron) of the Dirac cone is not coupled to the WSM surface state, while the other part is maximally coupled, see \eq{H_WTI^sur_2} above. 
For the lower, hole-like cone, using the parameters in \figr{surf_disp_a0}, this direction is 
$\phi_k=-\fr{\pi}{2}$, thus the negative $k_y$ axis with $k_x=0$. The dispersion relation is then $E=C + A k_y$
corresponding to the cyan line in \figr{surf_disp_a0} (a) which crosses the other two straight lines. 

Considering finite couplings $a_0\neq0$ or $b_1\neq0$ gives only quantitative differences in the dispersion relations (not shown). The Dirac point generation is unaffected, 
except for the special case where the Dirac point and Fermi arc cross only at $k_x=k_y=0$. As in this case the coupling for $b_1\neq0$ is absent in this point, no 
second Dirac point is generated. 

A perturbative calculation can provide some insight into both kinds of Dirac points. We take the surface Hamiltonian, \eq{H_WTI^sur}, 
and treat one band as a perturbation to the other two. 

For the shifted Dirac point one directly finds in 2nd order perturbation theory in the coupling
\beq \bsp{
	H_{D}^{1} & = \lt(\ba{cc} C & i\eta Ak_{-} \\
	-i\eta A^*k_{+} & C \ea\rt) \\ 
	& + \fr{1}{C-\ga \lt(M_0 + M_1 k_x^2\rt)-\eta v_yk_{y}} \lt(\ba{cc} \lt|H_{c}\rt|^2  & H_{c} \Ht_{c}^* \\
	\Ht_{c} H_{c}^* & |\Ht_{c}|^2 \ea\rt). \lbl{eq:H_shifted_dirac}
} \eeq
Evidently, a difference in the absolute values of the coupling between the Weyl system and the different spin species of the TI system will open a gap. 
In the limit of spin degeneracy, where $H_c=\Ht_{c}$, we insert the coupling from \eq{H_c}, expand \eq{H_shifted_dirac} for small momenta and find 
\beq \bsp{
	H_{D}^{1} & = \lt(\ba{cc} C & i\eta Ak_{-} \\
	-i\eta A^*k_{+} & C \ea\rt) \\
	& + \fr{\lt|a_0\rt|^2}{C-\ga M_0}\lt(\ba{cc} 1 & 1 \\ 1 & 1 \ea\rt) +\mathcal{O}\lt(k_\pm\rt)
	\lbl{eq:H_shifted_dirac_linearized}
} \eeq
corresponding to a Dirac cone shifted in energy and momentum by the coupling. For real $A$ the shift occurs 
in the $k_y$ direction as shown in \figr{surf_disp_a0}.

The creation of the second Dirac point can be understood from a similar calculation. The perturbative Hamiltonian for this Dirac point is given by
\bwt
\beq
	H_{D}^{2} = \lt(\ba{cc} C - \lt|A\rt| k_\Vert & \fr{1}{\sq{2}} H_{c} \lt(1-e^{i\phi_k^A}\rt) \\
	\fr{1}{\sq{2}} H_{c}^* \lt(1-e^{-i\phi_k^A}\rt)\ \  & \ga \lt(M_0 + M_1 k_x^2\rt)+\eta v_yk_{y} 
	- \lt|H_{c}\rt|^2 \fr{1+\cos\lt(\phi_k^A\rt)}{C + \lt|A\rt| k_\Vert - \ga \lt(M_0 + M_1 k_x^2\rt)-\eta v_yk_{y}}\ea\rt).
	\lbl{eq:H_created_dirac}
\eeq
\ewt
The off-diagonal elements vanish along 
the momentum direction $\phi_k=\phi_A + \eta\fr{\pi}{2}$. Thus, Weyl and Dirac surface states are uncoupled in one point. This point 
becomes the new Dirac point, and setting the diagonal elements of \eq{H_created_dirac} equal, this gives its precise value $k_D$. 
For the parameters used in \figr{surf_disp_a0}, the Dirac point $k_D$ is on the negative $k_y$ axis, with the corresponding Hamiltonian 
\beq
	H_{D}^{2} = \lt(\ba{cc} C + A \lt(k_D + k_y\rt) & \fr{1}{\sq{2}} k_x \lt(b_1 +i \fr{a_0}{k_D}\rt) \\
	\fr{1}{\sq{2}} k_x \lt(b_1 -i \fr{a_0}{k_D}\rt) & C + A \lt(k_D - k_y\rt) + 2f_{di}(k_x,k_y) \ea\rt)
	\lbl{eq:H_created_dirac_local}
\eeq
including the distortion $f_{di}(k_x,k_y)=\lt(A-v_y\rt) k_y+\fr{2b_1\lt(a_0 k_x-x_1 k_D k_y \rt)+A k_D k_y\lt(A-v_y\rt)}{C - \lt(A-v_y\rt)k_D - \ga M_0}$ 
and $k_D=\fr{v_y \lt(C-\ga M_0\rt)-\sq{2a_0^2 \lt(A^2-v_y^2-2b_1^2\rt)+\lt(A^2-2b_1^2\rt)\lt(C-\ga M_0\rt)^2}}{A^2-v_y^2-2b_1^2}$. The Dirac point is stable for 
any real combination of spin-symmetric couplings. A finite distortion $f_{di}\neq 0$ tilts the Dirac cone but does not open a gap. 

The number of Dirac points in the surface dispersion relation can be extended further by 
a coupling that changes sign as a function of $k_x$ and $k_y$, e.g. by setting $a_0>0$ and 
$a_2<0$ or a combination of $a_0\neq0$ and $b_1\neq0$. The positions in $k$-space where the coupling is zero and TI and WSM surface state 
intersect will then harbor additional Dirac points, see \figr{surf_disp_a0_a2}. 
\bfg%[!]
	\includegraphics[width=0.45\textwidth]{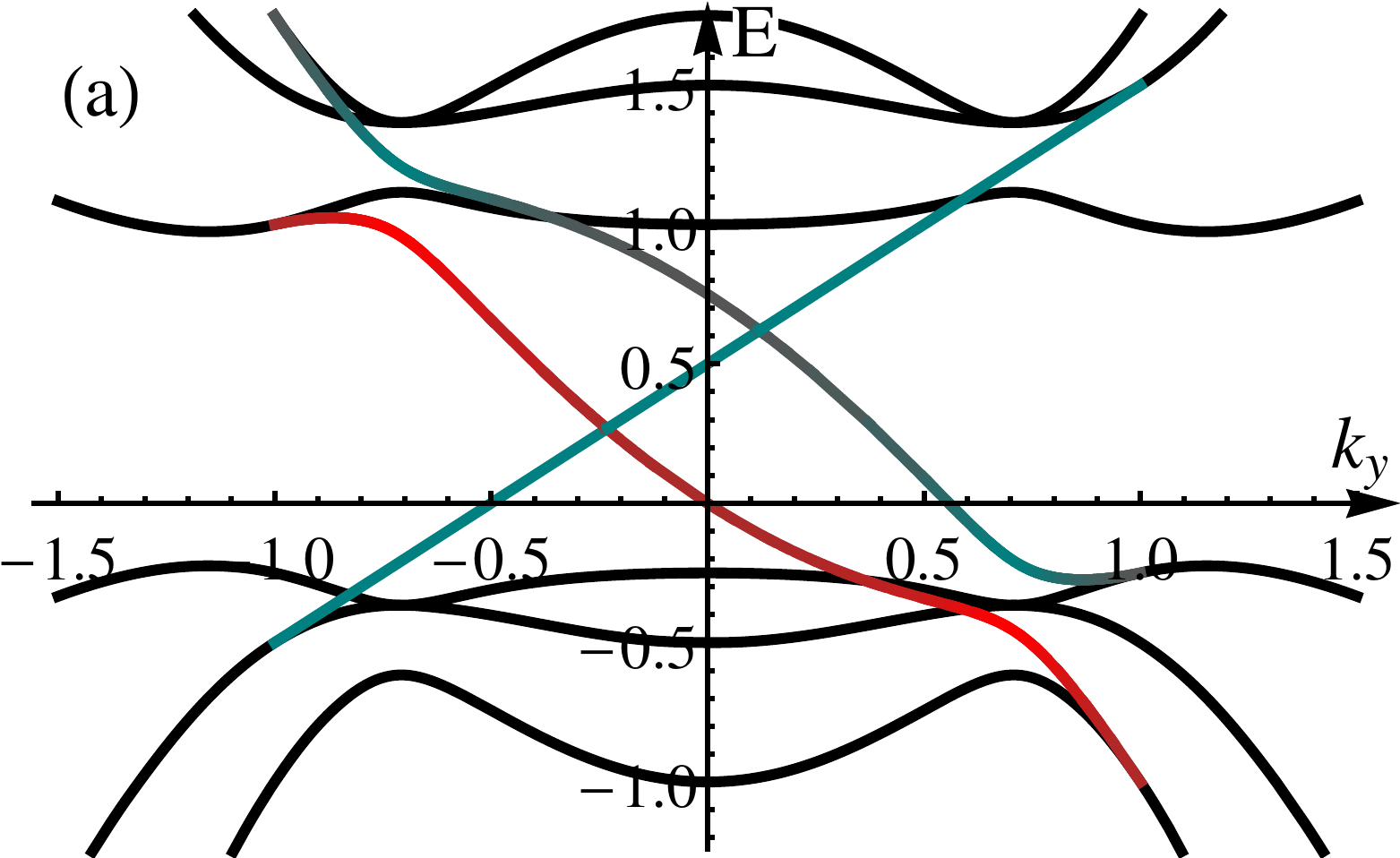}
	\includegraphics[width=0.4\textwidth]{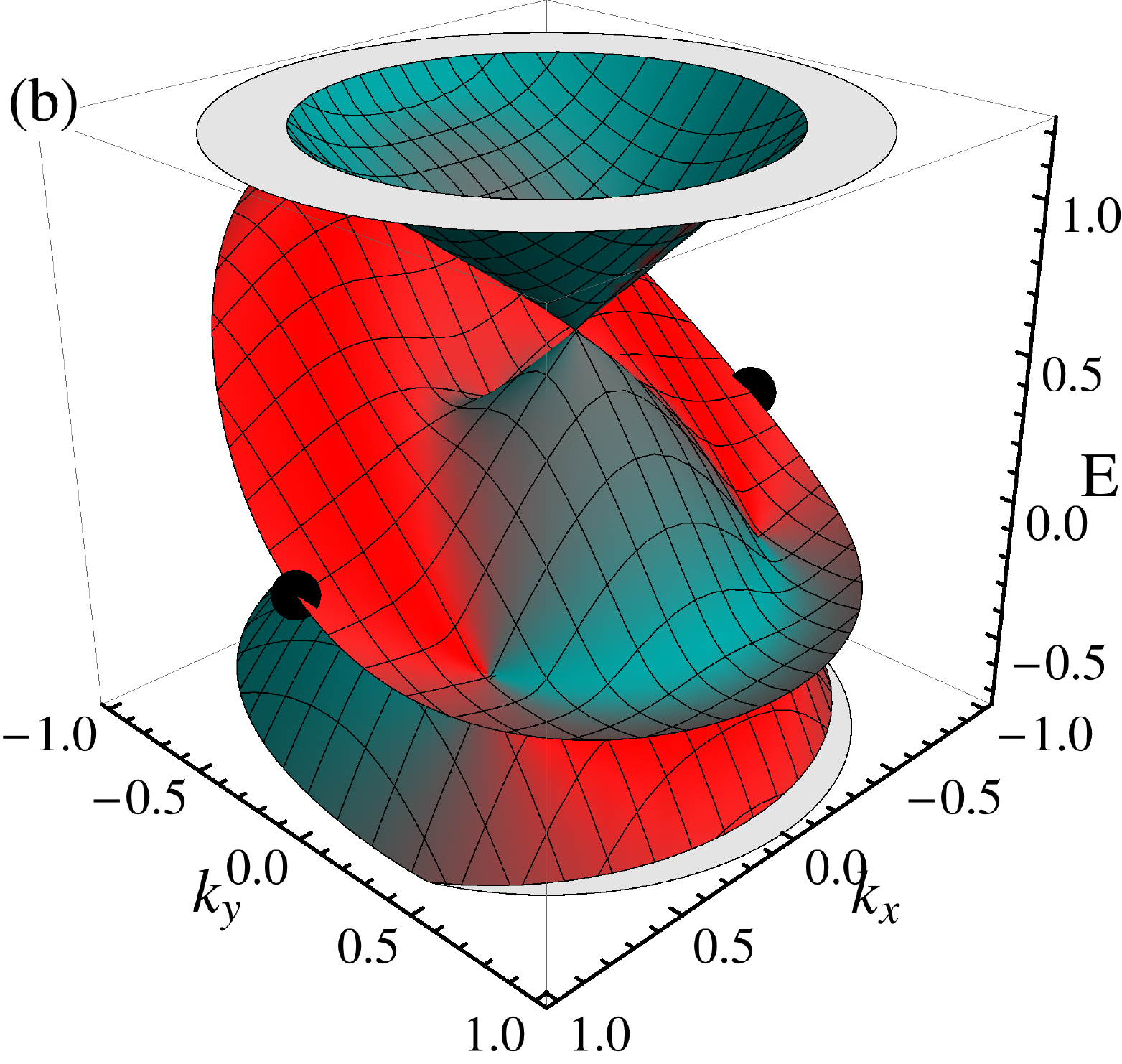}
	\caption{(a) Bulk and upper surface dispersion relations of the TI and WSM models with a coupling that changes sign. 
	Color code: Black lines for the bulk states, red for WSM character and cyan for the TI character of the surface states.
	(b) 3D plot of the surface dispersion relation. Four Dirac points are visible. 
	\\ 	
	Parameters: $C=\fr{1}{2}$, $M_0=-1$, $M_1=1$, $B=1$, $k_z=0$, $\ga=-\fr{1}{4}$, $A=1$, $v_y=1$, $a_0=\fr{1}{4}$, $a_2=-\fr{1}{2}$, $b_1=0$, $H_{c}=\Ht_{c}$.}
	\lbl{fig:surf_disp_a0_a2}
\efg

\subsection{Spin-asymmetric coupling: Creation of gaps \& spin polarization}

The spin-up and spin-down TI bands do not need to have the same coupling to the WSM. If the absolute values 
are different, $\lt|H_{c}\rt|\neq|\Ht_{c}|$, 
the Dirac points in the surface dispersion relation are gapped out, see \eq{H_shifted_dirac} and \figr{surf_disp_a0_asym}. 
\bfg[!ht]
	\includegraphics[width=0.45\textwidth]{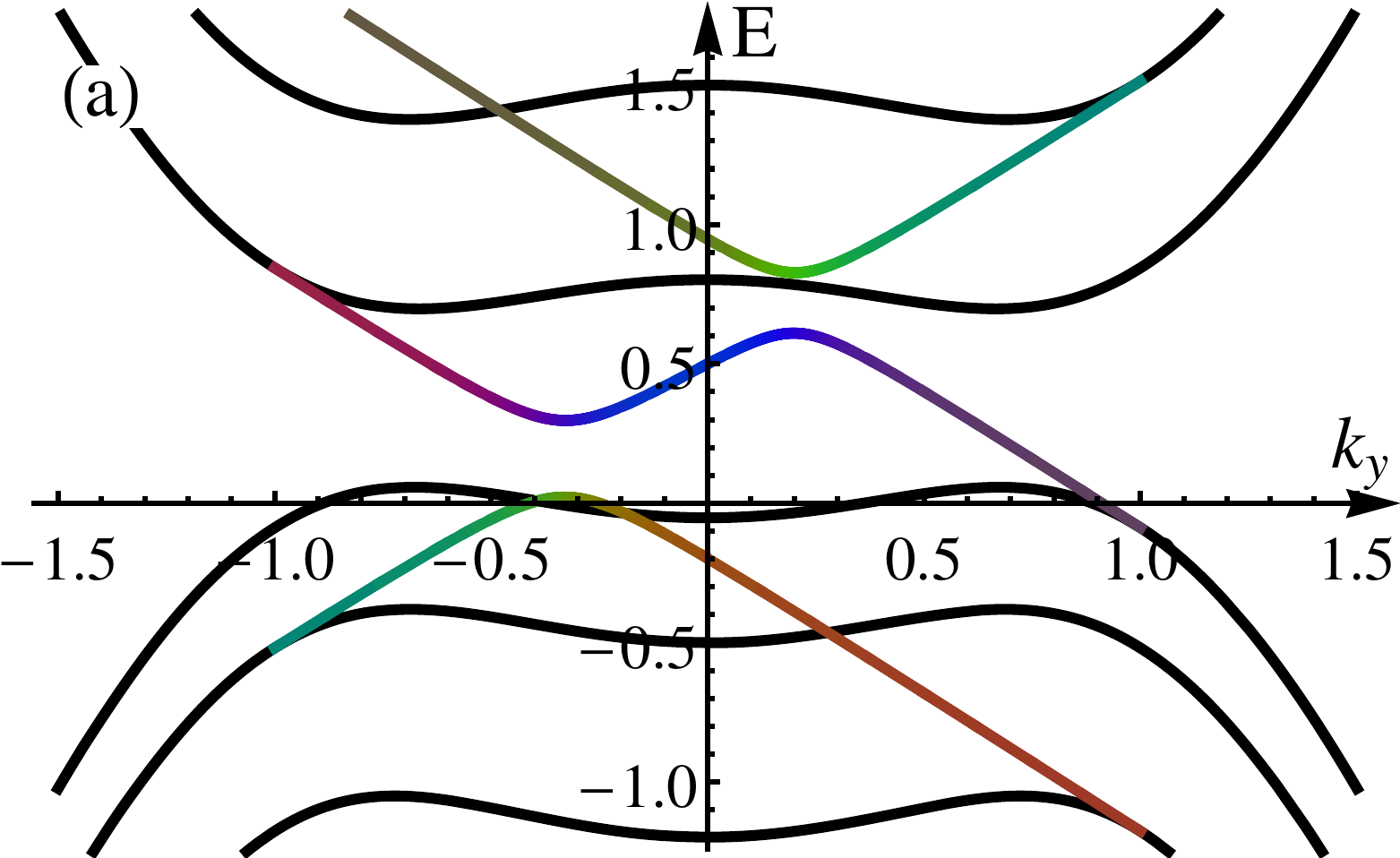}
	\includegraphics[width=0.45\textwidth]{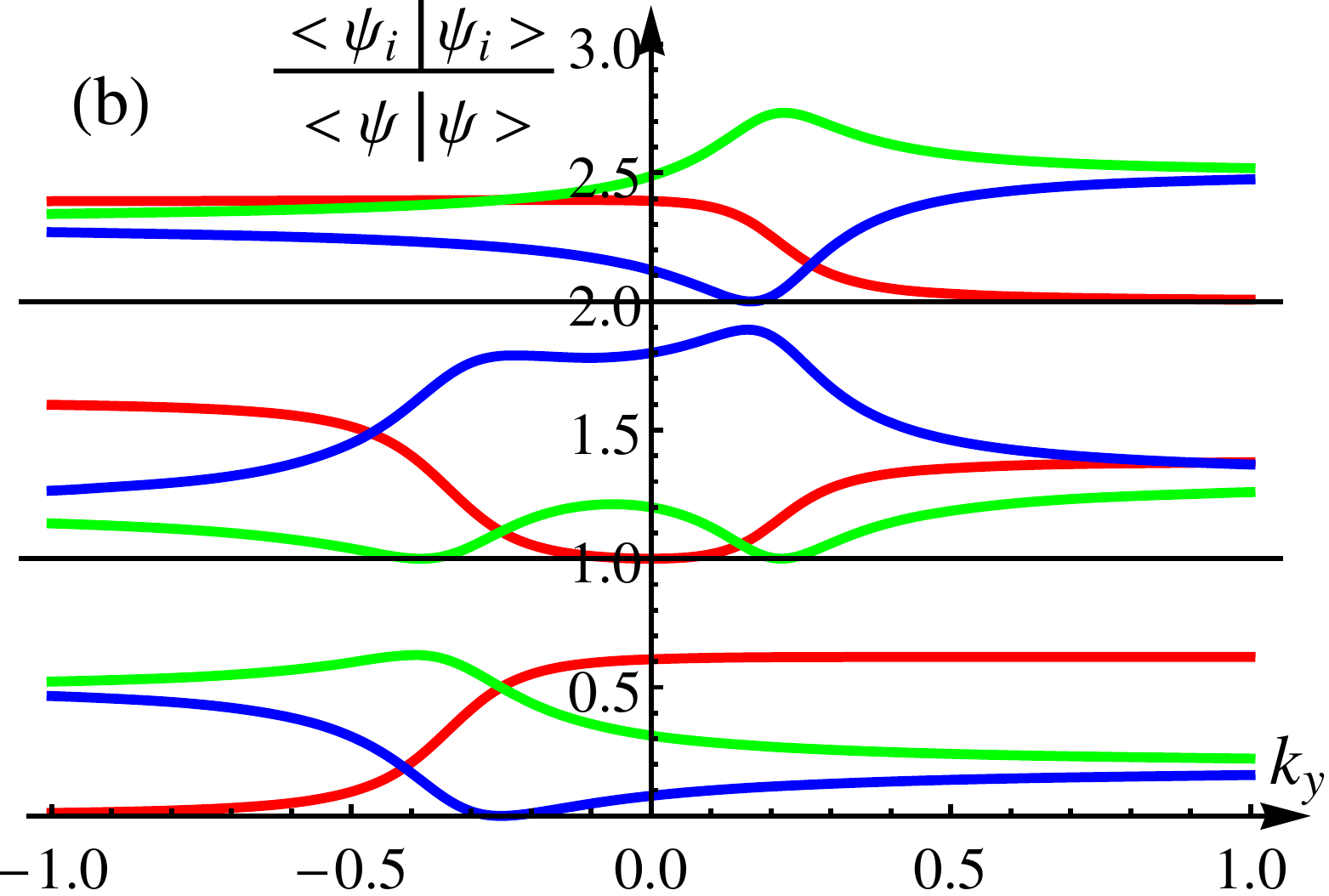}
	\includegraphics[width=0.4\textwidth]{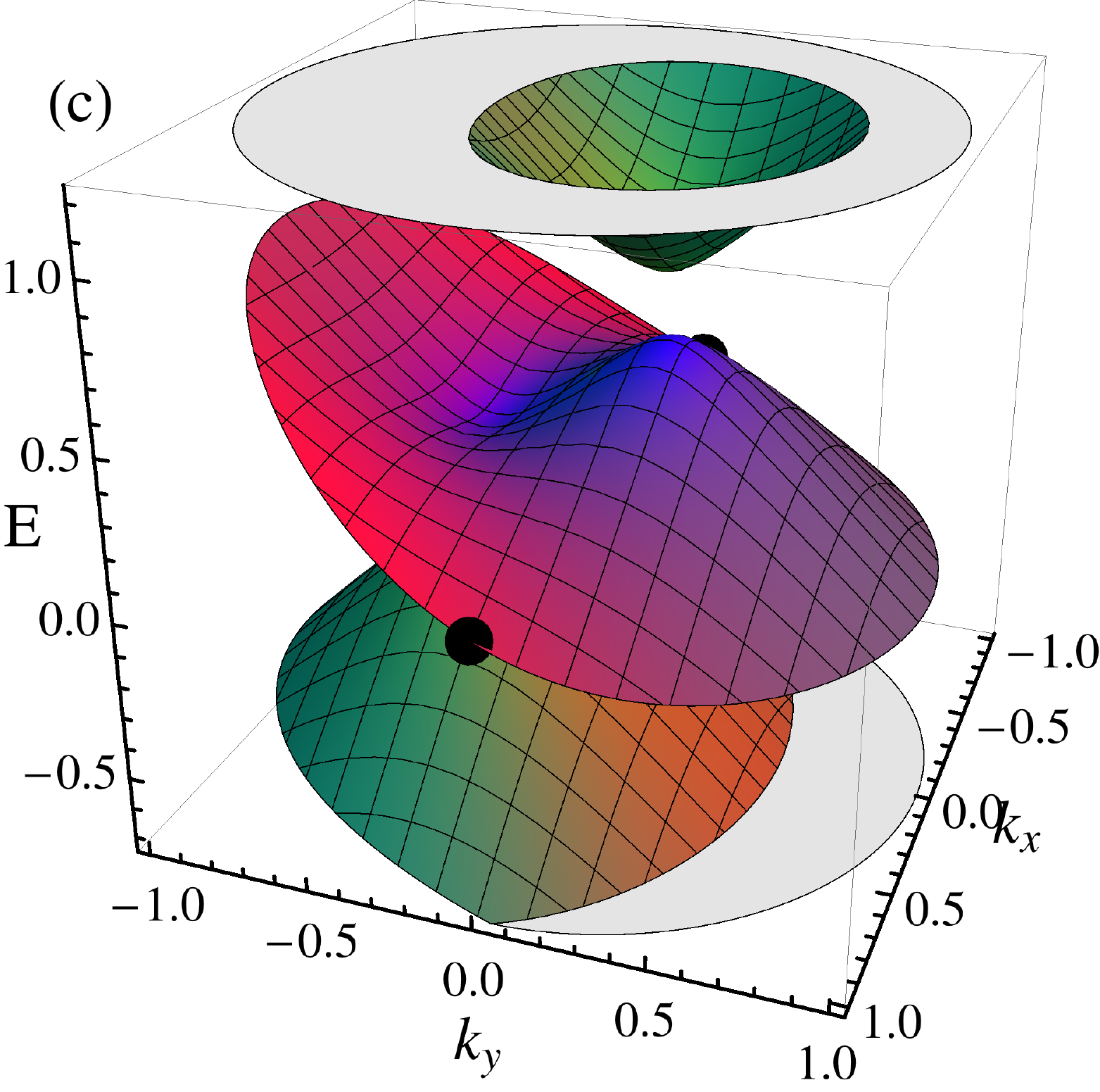}
	\caption{(a) Bulk and upper surface dispersion relations of the TI and WSM model with an spin-asymmetric coupling. 
	Color code: Black lines for the bulk states, red for WSM character and blue (green) for the TI spin up (down) character of the surface states.
	(b) Character of the three surface bands, shifted for clarity. 
	(c) 3D plot of the surface dispersion relation. All Dirac points are gapped.  \\ 
	Parameters: $C=\fr{1}{2}$, $M_0=-1$, $M_1=1$, $B=1$, $k_z=0$, $\ga=-\fr{1}{4}$, $A=1$, $v_y=1$, $a(k_\Vert)=\fr{1}{4}$, $\tilde{a}(k_\Vert)=\fr{2}{4}$, $b_1=\tilde{b}_1=0$.}
	\lbl{fig:surf_disp_a0_asym}
\efg
The bulk Weyl points are, however, unaffected. 
The resulting bands are partly spin polarized as shown in \figr{surf_disp_a0_asym}. The weaker coupled spin up
electrons form a band with the Weyl surface state at intermediate energies, while the stronger coupled spin down electrons are pushed into the upper and 
lower bands. Considering a finite $b_1\neq0$ instead of a $a_0$ coupling, only the lower Dirac point will split. 
As the upper one is located at $k_x=k_y=0$ for a pure momentum dependent coupling, the effective coupling between the WSM and TI surface states is 
zero here. 

\subsection{Phase-shifted coupling: Moving Dirac points, tilting dispersion relation}

Including complex coupling constants, this offers additional ways to alter the bulk and surface spectrum. 
In general, the dispersion relation will look much less symmetric compared to the previous, real couplings. 

Assuming $\Ht_{c}=H_{c}$, one can directly conclude from the Hamiltonian in \eq{H_WTI^sur_2} that a complex coupling $A=i$ 
will lead to two Dirac points lying on the $k_x$, rather than on the $k_y$ axis as discussed in \secr{real_sym_coup}. This is
confirmed in \figr{surf_disp_a0_Ai}. 
\bfg[!ht]
	\includegraphics[width=0.45\textwidth]{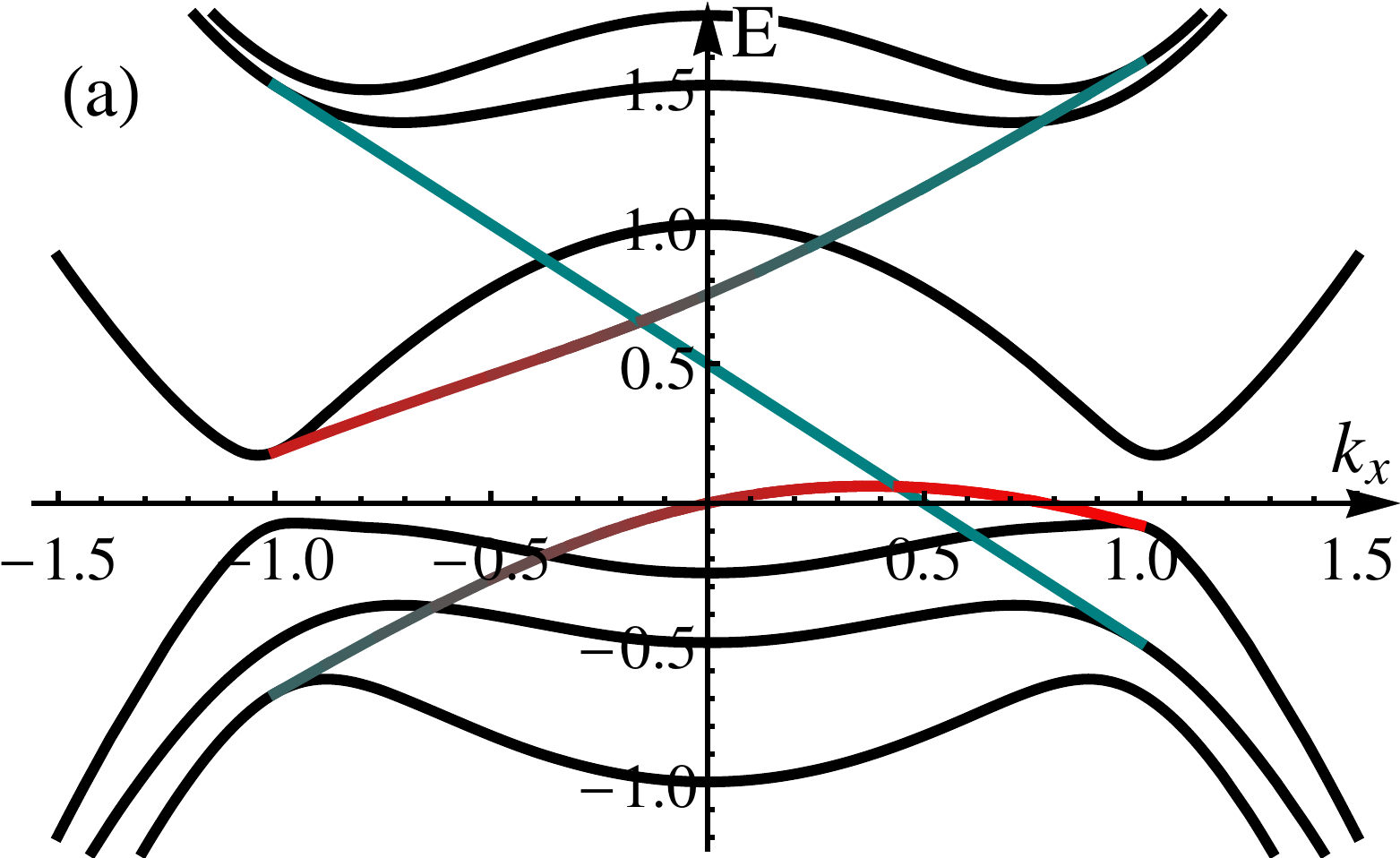}
	\includegraphics[width=0.45\textwidth]{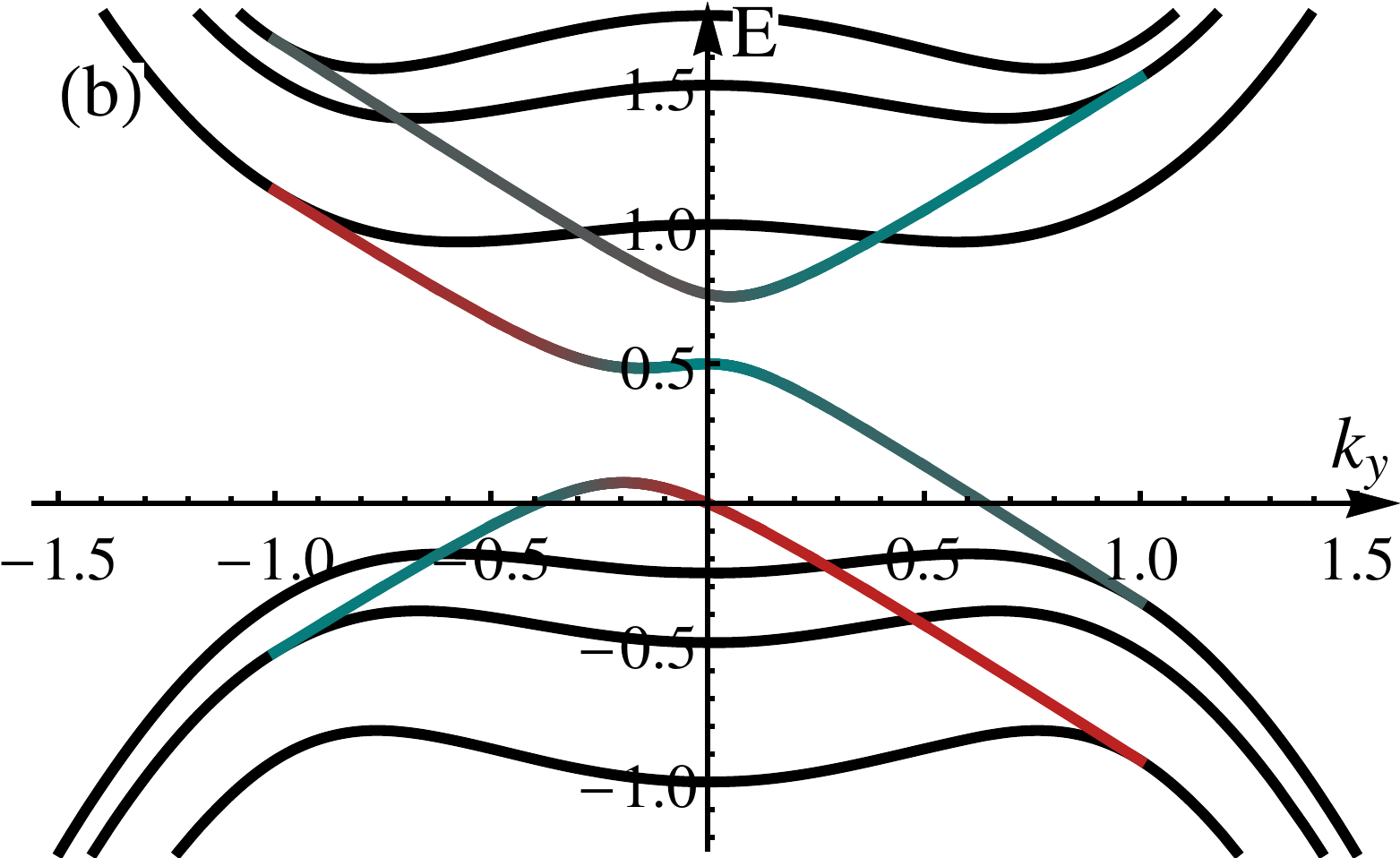}
	\includegraphics[width=0.4\textwidth]{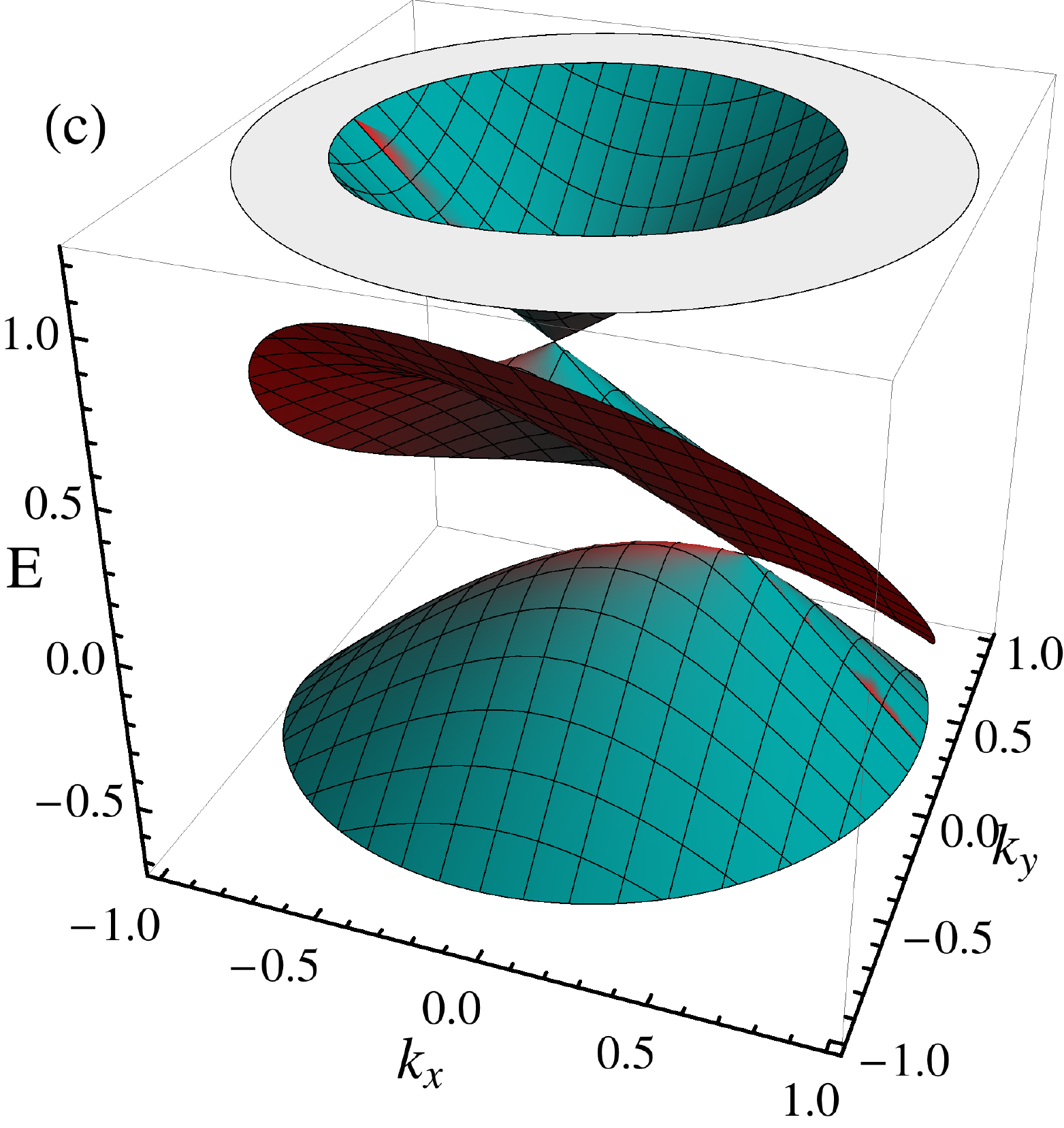}
	\caption{(a) and (b) Bulk and upper surface dispersion relations of the TI and WSM model with a complex coupling. 
	Color code: Black lines for the bulk states, red for WSM character and cyan for the TI character of the surface states.
	(c) 3D plot of the surface dispersion relation. Two Dirac points on the $k_x$ axis are visible. 
	 \\ 
	Parameters: $C=\fr{1}{2}$, $M_0=-1$, $M_1=1$, $B=1$, $k_z=0$, $\ga=-\fr{1}{4}$, $A=i$, $v_y=1$, $a(k_\Vert)=\fr{1}{4}$, $b_1=0$, $H_{c}=\Ht_{c}$.}
	\lbl{fig:surf_disp_a0_Ai}
\efg
One also sees that the bulk Weyl points lie not on the $k_x$ axis, but are rotated by the complex coupling. Yet
the rotation is much smaller than the $\pi/2$ rotation of the surface Dirac points. 

The same effect is obtained by a complex phase difference between the couplings $H_{c}$ and $\Ht_{c}$. It 
can even undo the rotation induced by $A=i$. 
Note also that in the spin-symmetric case, already for real and finite $a(k_\Vert)$ and $b_1$ the Weyl points are rotated away from the $k_x$ axis. 
Here the effective coupling is complex, with a phase changing with $k_\pm$. Supplementing this with a complex $a(k_\Vert)$, this can 
again lead to points where the effective coupling is zero, resulting in additional Dirac points like in \secr{real_sym_coup}.

\section{Experimental Realization}\lbl{sec:experiment}

We propose two ways to realize the physics of hybrid TI and WSM phases in an experimental setup. First, a material that naturally is in this combined phase will have corresponding surface states, as depicted in \figr{exp_realization} (a). 
\bfg
	\includegraphics[width=0.4\textwidth]{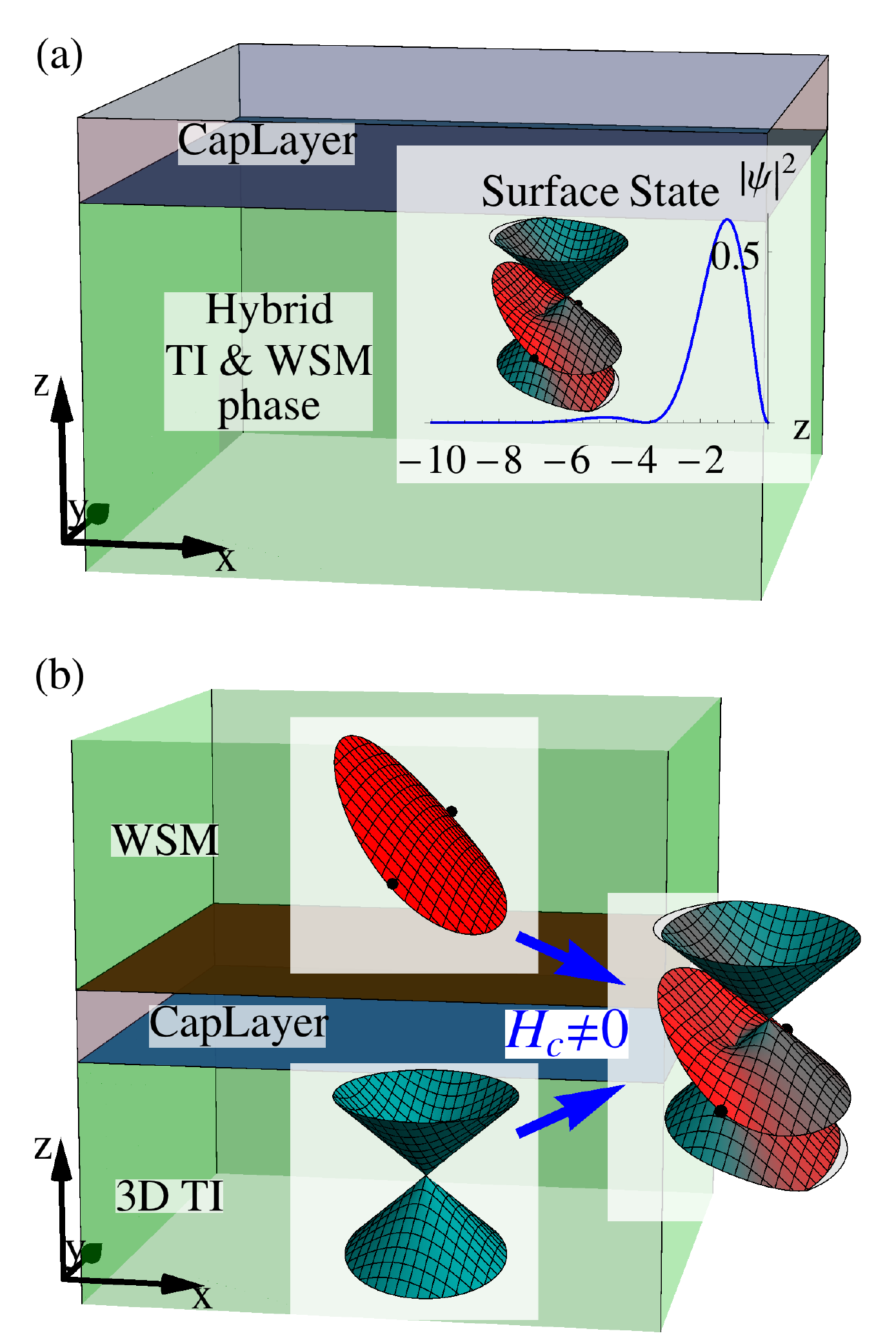}
	\caption{Possible experimental realizations: (a) Bulk materials being in the combined 3D TI and WSM phase will naturally have hybrid
	surface states. (b) A heterostructure where TI and WSM phases are adjacent to each other
	will exhibit hybrid surface states for finite coupling $H_C\neq0$, provided e.g. by Coulomb interaction or tunneling. }
	\lbl{fig:exp_realization}
\efg
Compressively strained HgTe is a candidate material for this phase: the compressive strain pushes the $\Ga_8$ bands against each other creating Weyl points\cite{Ruan2016}, in addition to the prevailing topological band inversion between the $\Ga_8$ and $\Ga_6$ bands. A difference to our calculation is the preserved time-reversal symmetry, leading to eight Weyl points in HgTe instead of two. 
However, if HgTe is doped with Mn the number of Weyl points could be reduced by a (partial) magnetic ordering. 

The second realization consists of a WSM in contact with a 3D TI, possibly separated by a thin buffer layer as depicted 
in \figr{exp_realization} (b). This should lead to a hybrid surface state at the joint boundary. 
The finite coupling $H_C$ could be provided by tunneling or Coulomb interaction. While this surface state is not exactly of the form of the ansatz in 
\eq{simple_ansatz_WTI}, the surface Hamiltonian, \eq{H_WTI^sur}, should still be valid with the modification $\eta=\eta_{TI}=-\eta_{W}$. 
As several proposals of TRS-broken WSM with two Weyl points are based on magnetically doped 3D TI materials~\cite{Burkov2011,Cho2011,Xu2011,Bulmash2014}, 
the fabrication of the described hybrid system should be technically feasible. 

\section{Conclusion \& Outlook}\lbl{sec:conclusion}

We have analyzed a hybrid system composed of a 3D TI coupled to an inversion symmetric, TRS-broken WSM. 
In the spirit of a tunnel coupling approach between the two topological phases, the use of a simplified ansatz made it possible to find an analytical solution for the surface states. 
The resulting surface Hamiltonian, \eq{H_WTI^sur}, is a major result of this paper. 
The dispersion relation of the hybrid system shows different phenomena depending on the assumed coupling between WSM and TI. 
Preserved spin symmetry e.g. leads to the creation of additional Dirac points in the surface dispersion relation. Breaking of spin symmetry 
on the other hand, this opens gaps and induces spin polarization in the former Dirac surface cone. As an experimental realization we have 
presented both strained HgTe, which might naturally be in the discussed hybrid phase, and a heterostructure of TI and WSM. In the latter case, the 
joint boundary would harbor the interesting hybrid surface state. 

There are several directions how to proceed with this research. Looking for measurable consequences, e.g. in transport or spectroscopy, of the 
new hybrid surface states should be the most immediate one. We expect, for instance, that different Dirac points 
will give rise to different minima in the conductivity, similar to the graphene case~\cite{Katsnelson2006, Tworzydlo2006}. 
An extension to time-reversal symmetric WSM is another one. For this, one 
should use a 4x4 Hamiltonian for the WSM, which offers the possibility of more involved Fermi arcs on the surface, e.g. including spin polarization along the arcs~\cite{Lv2015c,Xu2016b}. 
TaIrTe$_4$~\cite{Koepernik2016,Belopolski2016} with its four Weyl points could be a candidate material for a hybrid system of this kind. 

\begin{acknowledgments}
We thank C. Br\"une, E.~M. Hankiewicz, J.~B. Mayer and M. Kharitonov for interesting discussions. 
 We acknowledge financial support by the DFG (SPP 1666 and SFB 1170 "ToCoTronics"), the Helmholtz Foundation (VITI) as well as 
the ENB Graduate school on "Topological Insulators".
\end{acknowledgments}

%\newpage

\appendix

\section{Hardwall boundary condition 2x2}\lbl{sec:app_boundary}
In this section, we recap a simple method for calculating exponentially localized boundary states of a 2x2 Hamiltonian following 
Ref.~\ocite{Liu2010} and references therein. We introduce hardwall boundary conditions on a half space $z\leq0$ or $z\geq0$. Thus, 
the surface state is localized at $z=0$ and decays either in direction $z\ra-\iy$ (upper surface) or $z\ra+\iy$ (lower surface). The state should fulfill the eigenvalue equation 
\beq
	H\Psi\lt(z\rt)=E\Psi\lt(z\rt)
	\lbl{eq:eigeneq}
\eeq
with $H=\lt[h_4\lt(k_\Vert^2+k_z^2\rt)+h_3\rt]\tau_3+h_2k_z\tau_2+h_1\lt(k_{\pm}\rt)\tau_1+h_0\tau_0$ and $h_j$ being real constants or functions of $k_\pm$. 
The Hamiltonian can represent a topological insulator or Weyl semimetal depending on the chosen $h_j$. 

The general ansatz for the eigenstate is 
\beq
	\Psi_g\lt(z\rt)=\sum_{j\in\{1,2\}} a_je^{ik_{z,j}z}\psi\lt(k_{\pm},k_{z,j}\rt), \lbl{eq:gen_ansatz_app}
\eeq
which could be used to solve for the surface states of \eq{eigeneq} in the usual manner. 
Due to the specific structure of our Hamiltonian, we can choose a simplified version of the ansatz, given by 
\beq
	\Psi\lt(z\rt)=\lt(e^{ik_{z,1}z}-e^{ik_{z,2}z}\rt)\psi\lt(k_{\pm}\rt). \lbl{eq:simple_ansatz}
\eeq
The relative sign ensures that the wave function vanishes at $z=0$. The ansatz offers the possibility to separate \eq{eigeneq} into two parts
\bal{
	\lt[h_1\lt(k_{\pm}\rt)\tau_1+h_0\tau_0\rt]\Psi\lt(z\rt) &= E\tau_0\Psi\lt(z\rt) \lbl{eq:eigeneq_disp}\\
	\{[h_4(k_\Vert^2+k_z^2)+h_3]\tau_3+h_2k_z\tau_2\}\Psi\lt(z\rt) &= 0. \lbl{eq:eigeneq_k_z}
}
\eq{eigeneq_disp} is independent of $k_z$ and can be solved for the surface dispersion relation $E$, while the solution of \eq{eigeneq_k_z} defines the two quantized values of $k_z$ needed for the surface eigenstate. 

Following this procedure, $\psi\lt(k_{\pm}\rt)=f\lt(k_{\pm}\rt)\psi_\pm$ 
is taken to be proportional to the eigenstate of the $\tau_1$ Pauli matrix, 
$\tau_1\psi_{\pm}=\pm\psi_\pm$ and 
\beq
	\psi_\pm=\fr{1}{\sq{2}}\lt(\ba{c} 1 \\ \pm1 \ea\rt)
\eeq 
with $f\lt(k_{\pm}\rt)=1$. Using $\tau_2\psi_{\pm}=\mp i\psi_{\mp}$ 
and $\tau_3\psi_{\pm}=\psi_{\mp}$, \eq{eigeneq_k_z} reduces to the quadratic equation
\beq
	h_4\lt(k_\Vert^2+k_z^2\rt)+h_3-\eta i h_2k_z = 0
\eeq
with $\eta=\pm$ the sign inherited from $\psi_\pm$. Solving for $k_z$, we find the two solutions
\beq
	ik_{z,\substack{1\\2}}=\fr{1}{2h_4} \lt[-\eta h_2 \pm\sq{4h_4\lt(h_3+h_4 k_\Vert^2\rt)+h_2^2}\rt]. \lbl{eq:ik_z_2d}
\eeq
In order to obtain a wave function, exponentially decaying of the form of \eq{simple_ansatz}, both $ik_{z,\substack{1\\2}}$ need a real part of the same sign. 
For real $h_j$, this gives us the existence condition
\beq
	h_4\lt(h_3+h_4 k_\Vert^2\rt)<0. \lbl{eq:exist_cond_gen}
\eeq
Depending on the sign of $h_2/h_4$ and the direction in which the wave function should decay, $z\ra+\iy$ or $z\ra-\iy$, 
one chooses the corresponding eigenstate $\psi_\pm$, fixing 
\beq
	\eta=- \sgn\lt(\fr{h_2}{h_4}\rt),\ {\rm top};\ \eta=\sgn\lt(\fr{h_2}{h_4}\rt),\ {\rm bottom}.
\eeq 
The surface dispersion relations and wave functions are then given by 
\beq
	E^{sur} = h_0 + \eta h_1\lt(k_\pm\rt);\ 
	\Psi\lt(z\rt)=\lt(e^{ik_{z,1}z}-e^{ik_{z,2}z}\rt)\psi_{\eta}.
\eeq
The localization length is  $l_c=\max\lt\{\lt|1/\Re\lt(ik_{z,\substack{1\\2}}\rt)\rt|\rt\}$. 

The surface solution described in this section fulfills the eigenvalue \eq{eigeneq}
and is thus a valid, non-perturbative eigenstate of the Hamiltonian. Calculating the surface state with the general ansatz (\ref{eq:gen_ansatz_app}), 
this gives the same dispersion relation as for the simplified ansatz (\ref{eq:simple_ansatz}) for the TI model in \secr{model_TI}. 

\section{Numerical validation of approximate solution method}\lbl{sec:app_numerics}
In this article we use an analytical method to calculate the localized boundary states, described in App.~\ref{sec:app_boundary}. It requires 
certain restrictions on the parameters of the coupled TI-Weyl Hamiltonian, such as the same localization length for both TI and Weyl phase and half of 
the symmetry allowed couplings to be zero, see \secr{coupled_sys}. 

These constraints might seem quite restrictive. In order to proof the general applicability of our results, we have checked numerically that the neglected couplings 
have no qualitative effect on the surface band structure if kept reasonably small. The same is true for variations that 
alter the localization lengths of the subsystems. The numerical method is similar to the analytical approach: We solve for exponentially localized 
surface wave functions on the half space $z\leq0$ or $z\geq0$ with hard-wall boundary condition at $z=0$. The differences to the approximate solution is the use 
of the full ansatz for the wave function, i.e. 
\beq
	\Psi_g\lt(z\rt)=\sum_{j\in\{1,\ldots, 6\}} a_je^{ik_{z,j}z}\psi\lt(k_{\pm},k_{z,j}\rt). \lbl{eq:gen_ansatz_app_2}
\eeq

As an example, we take the case of the generation of the second Dirac point, discussed in \secr{real_sym_coup} and depicted 
in \figr{surf_disp_a0}. Besides the finite $a_0=\fr{1}{4}$, we add an additional coupling $d_0=\fr{1}{8}$ or change the localization length of the Weyl 
Hamiltonian by setting $v_z=\fr{3}{4}\neq B=1$ and $t=\fr{5}{4}\neq M_1=1$. The latter choice leads then to differing localization lengths of the separate systems of
\bal{
	ik_{z,TI} &= \fr{1}{2} \lt[\fr{B}{M_1} \pm\sq{4 \lt(k_\Vert^2 -1 \rt)+\lt(\fr{B}{M_1}\rt)^2}\rt], \\
	ik_{z,WSM} &= \fr{1}{2} \lt[\fr{v_z}{t} \pm\sq{4 \lt(k_\Vert^2 -1 \rt)+\lt(\fr{v_z}{t}\rt)^2}\rt]. 
}
The resultant dispersion relations are shown in \figr{numerics}, depicted by blue dots. 
\bfg
	\includegraphics[width=0.45\textwidth]{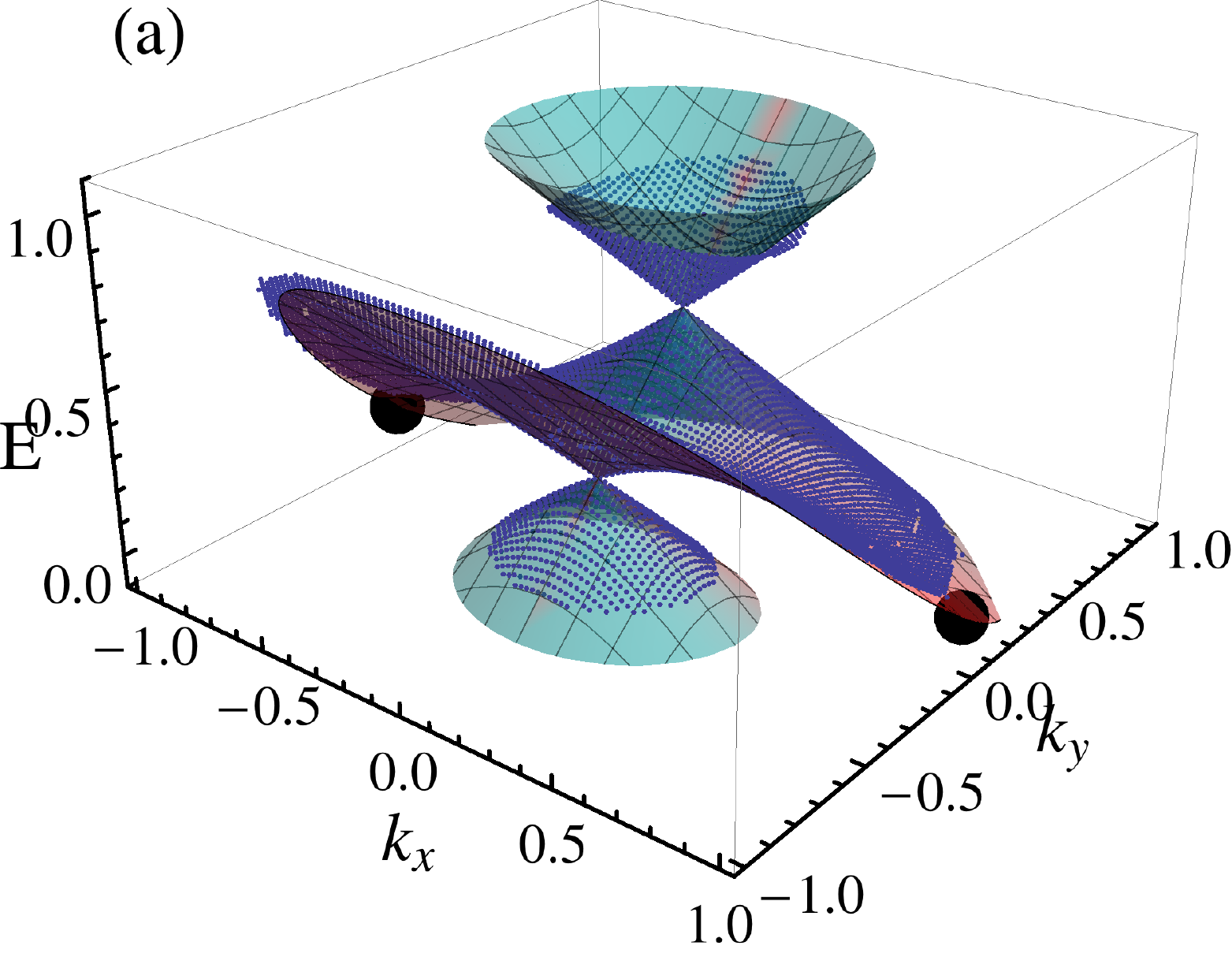}
	\includegraphics[width=0.45\textwidth]{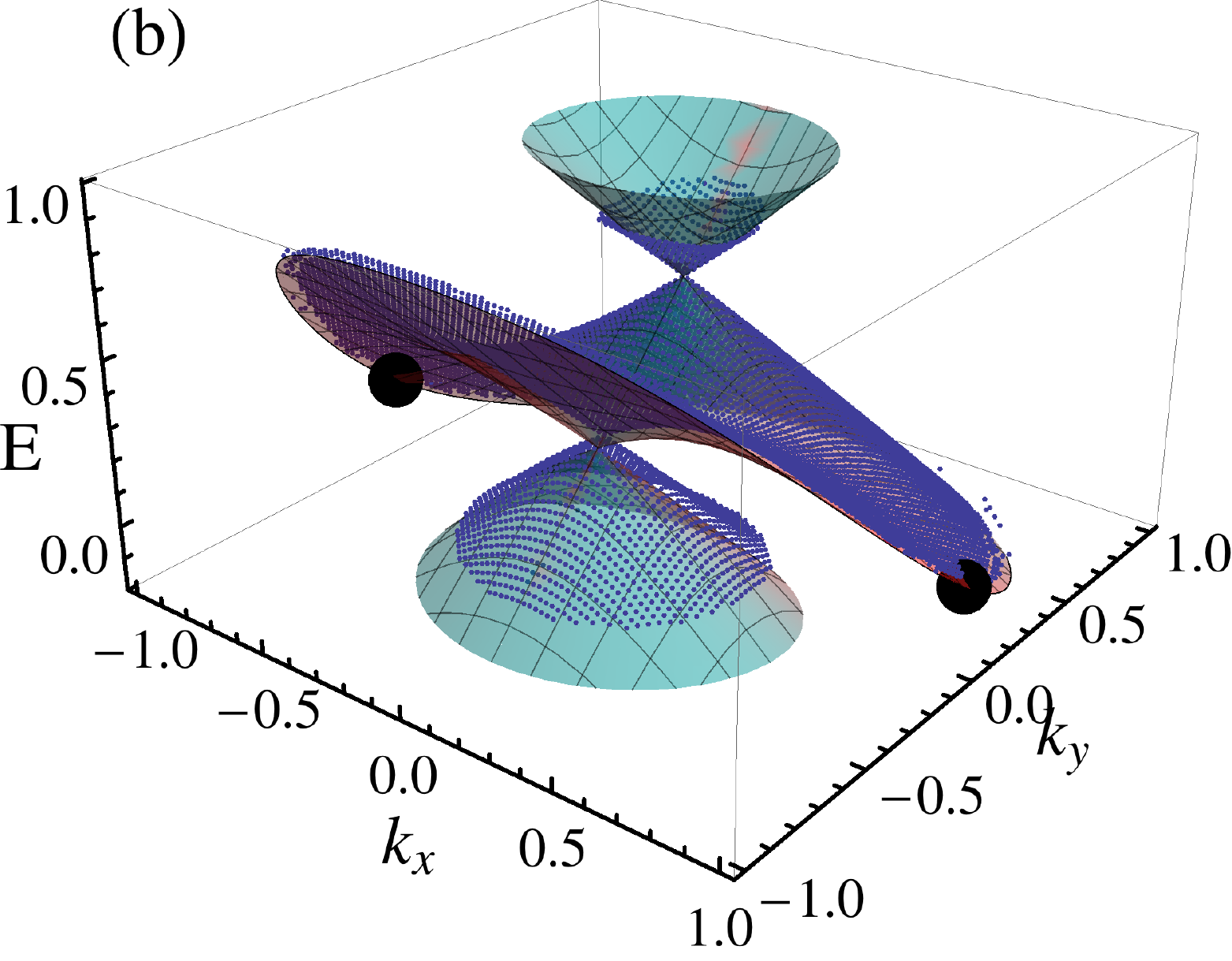}
	\caption{Dispersion of the upper surface of the combined TI and WSM models. The continuous surface is the analytical solution from \figr{surf_disp_a0}, the blue dots represent the full numerical solution. In (a) the additional coupling $d_0=\fr{1}{8}$ was considered in the numerical solution, in (b) the altered parameters $v_z=\fr{3}{4}\neq B=1$ and $t=\fr{5}{4}\neq M_1=1$. The large black dots denote the position of the Weyl notes of the analytical solution.}
	\lbl{fig:numerics}
\efg
The analytical solution for the $a_0=\fr{1}{4}$ coupling alone is displayed as a continuous surface. 

First we notice that the numerical and analytical solution agree very well and show no qualitative difference. The added coupling $d_0=\fr{1}{8}$ 
in \figr{numerics} (a) has almost no effect, the same was found for a finite $c_1=\fr{1}{8}$. The changed localization length in 
\figr{numerics} (b) shifts a bit the lower Dirac point, but does not open a gap. The major difference between the analytical and numerical solutions is 
the restriction of the surface solution to energies and momenta where no bulk state exists. This becomes especially clear for the 
upper and lower parts of the Dirac cone in \figr{numerics}, and is due to hybridization between the bulk and surface states. It 
prevents the existence of purely exponentially localized surface wave functions in this parameter range.  

We conclude that the physical results and conclusions of this paper are valid beyond the restrictions on allowed couplings and localization lengths which are necessary to 
keep the analytical form of the equations simple.


\begin{thebibliography}{24}
% 3D TI Reviews
\bibitem{Hasan2010} M.~Z. Hasan and C.~L. Kane, Rev.~Mod.~Phys.~{\bf 82}, 3045 (2010). 

\bibitem{Qi2011} X.-L. Qi and S.-C. Zhang, Rev.~Mod.~Phys.~{\bf 83}, 1057 (2011).

\bibitem{Ando2013} {Y. Ando, J.~Phys.~Soc.~Jpn.~{\bf 82}, 102001 (2013).}
% WSM Reviews
\bibitem{Hosur2013} P. Hosur and X. Qi, C.~R.~Phys.~{\bf 14}, 857-870 (2013).

\bibitem{Turner2013} A.~M. Turner and A. Vishwanath, Contemporary Concepts of Condensed Matter Science~{\bf 6}, 293-324 (2013).
%WSM Reviews experimental
\bibitem{Jia2016} S. Jia, S.-Y. Xu and M.~Z. Hasan, Nat.~Mat.~{\bf 15}, 1140 (2016).

\bibitem{Yan2017} B. Yan and C. Felser, Annu.~Rev.~Condens.~Matter~Phys.~{\bf 8}, 1-19 (2017).
% Nielsens
\bibitem{Nielsen1981} H.~B. Nielsen and M. Ninomiya, Phys.~Lett.~B~{\bf 105}, 219 (1981).

\bibitem{Nielsen1983} H.~B. Nielsen and M. Ninomiya, Phys.~Lett.~B~{\bf 130}, 389 (1983). 

\bibitem{Murakami2007} S. Murakami, New Journal of Physics~{\bf 9}, 356 (2007).
% YIrO prediction
\bibitem{Wan2011} X. Wan, A.~M. Turner, A. Vishwanath and S.~Y. Savrasov, Phys.~Rev.~B~{\bf 83}, 205101 (2011). 
% TaAs prediction
\bibitem{Weng2015} H. Weng, C. Fang, Z. Fang, B.~A. Bernevig and X. Dai, Phys.~Rev.~X~{\bf 5}, 011029 (2015). 

\bibitem{Huang2015} S.-M. Huang, S.-Y. Xu, I. Belopolski, C.-C. Lee, G. Chang, B. Wang, N. Alidoust, G. Bian, M. Neupane, C. Zhang, S. Jia, A. Bansil, H. Lin and M.~Z. Hasan, Nat.~Commun.~{\bf 6}, 7373 (2015).
% NbAs
\bibitem{Xu2015b} S.-Y. Xu, N. Alidoust, I. Belopolski, Z. Yuan, G. Bian, T.-R. Chang, H. Zheng, V.~N. Strocov,	D.~S. Sanchez, G. Chang, C. Zhang, D. Mou, Y. Wu, L. Huang, C.-C. Lee, S.-M. Huang, B. Wang, A. Bansil, H.-T. Jeng, T. Neupert, A. Kaminski, H. Lin, S. Jia and M.~Z. Hasan, Nat.~Phys.~{\bf 11}, 748 (2015).
% Photonic Crystal
\bibitem{Lu2015} L. Lu, Z.Wang, D. Ye, L. Ran, L. Fu, J.~D. Joannopoulos and M. Solja\v{c}i\'{c}, Science~{\bf 349}, 622 (2015).
% TaAs papers
\bibitem{Xu2015} S.-Y. Xu, I. Belopolski, N. Alidoust, M. Neupane, G. Bian, C. Zhang, R. Sankar, G. Chang, Z. Yuan, C.-C. Lee, S.- M. Huang, H. Zheng, J. Ma, D.~S. Sanchez, B. Wang, A. Bansil, F. Chou, P.~P. Shibayev, H. Lin, S. Jia and M.~Z. Hasan, Science~{\bf 349}, 613 (2015). 

\bibitem{Lv2015} B.~Q. Lv, H.~M. Weng, B.~B. Fu, X. P. Wang, H. Miao, J. Ma, P. Richard, X.~C. Huang, L. X. Zhao, G.~F. Chen, Z. Fang, X. Dai, T. Qian and H. Ding, Phys.~Rev.~X~{\bf 5}, 031013 (2015). 

\bibitem{Lv2015b} B.~Q. Lv,	N. Xu, H.~M. Weng, J.~Z. Ma, P. Richard, X.~C. Huang, L.~X. Zhao, G.~F. Chen, C.~E. Matt, F. Bisti, V.~N. Strocov, J. Mesot, Z. Fang, X. Dai, T. Qian, M. Shi and H. Ding, Nat.~Phys.~{\bf 11}, 724 (2015).

\bibitem{Yang2015} L.~X. Yang, Z.~K. Liu, Y. Sun, H. Peng, H.~F. Yang, T. Zhang, B. Zhou, Y. Zhang, Y.~F. Guo, M. Rahn, D. Prabhakaran, Z. Hussain, S.-K. Mo, C. Felser, B. Yan and Y.~L. Chen, Nat.~Phys.~{\bf 11}, 728 (2015).
% MoTe2 experiment
\bibitem{Huang2016} L. Huang, T.~M. McCormick, M. Ochi, Z. Zhao, M.-T. Suzuki, R. Arita, Y. Wu, D. Mou, H. Cao, J. Yan, N. Trivedi and A. Kaminski, Nat.~Mat.~{\bf 15}, 1155 (2016).

\bibitem{Deng2016} K. Deng, G. Wan, P. Deng, K. Zhang, S. Ding, E. Wang, M. Yan, H. Huang, H. Zhang, Z. Xu, J. Denlinger, A. Fedorov, H. Yang, W. Duan, H. Yao, Y. Wu, S. Fan, H. Zhang, X. Chen and S. Zhou, Nat.~Phys.~{\bf 12}, 1105 (2016).

\bibitem{Tamai2016} A. Tamai, Q.~S. Wu, I. Cucchi, F.~Y. Bruno, S. Ricc\`{o}, T.~K. Kim, M. Hoesch, C. Barreteau, E. Giannini, C. Besnard, A.~A. Soluyanov and F. Baumberger, Phys.~Rev.~X~{\bf 6}, 031021 (2016).

\bibitem{Jiang2016} J. Jiang, Z.~K. Liu, Y. Sun, H. F. Yang, R. Rajamathi, Y.~P. Qi, L.~X. Yang, C. Chen, H. Peng, C.-C. Hwang, S.~Z. Sun, S.-K. Mo, I. Vobornik, J. Fujii, S.~S.~P. Parkin, C. Felser, B.~H. Yan and Y.~L. Chen, Nat.~Commun.~{\bf 8}, 13973 (2017).

\bibitem{Xu2016} N. Xu, Z.~J. Wang, A.~P. Weber, A. Magrez, P. Bugnon, H. Berger, C.~E. Matt, J.~Z. Ma, B.~B. Fu, B.~Q. Lv, N.~C. Plumb, M. Radovic, E. Pomjakushina, K. Conder, T. Qian, J.~H. Dil, J. Mesot, H. Ding and M. Shi, arXiv:1604.02116 (2016).
% MoTe2 prediction
\bibitem{Sun2015} Y. Sun, S.-C. Wu, M.~N. Ali, C. Felser and B. Yan, Phys.~Rev.~B~{\bf 92}, 161107(R) (2015).
% HgTe prediction
\bibitem{Ruan2016} J. Ruan, S.-K. Jian, H. Yao, H. Zhang, S.-C. Zhang and D. Xing, Nat.~Comm.~{\bf 7}, 11136 (2016).
% Chalcopyrites prediction 
\bibitem{Ruan2016b} J. Ruan, S.-K. Jian, D. Zhang, H. Yao, H. Zhang, S.-C. Zhang and D. Xing, Phys.~Rev.~Lett.~{\bf 116}, 226801 (2016).
% Ta3S2 prediction
\bibitem{Chang2015} G. Chang, S.-Y. Xu, D. S. Sanchez, S.-M. Huang, C.-C. Lee, T.-R. Chang, H. Zheng, G. Bian, I. Belopolski, N. Alidoust, H.-T. Jeng, A. Bansil, H. Lin and M. Z. Hasan, 
Sci.~Adv.~2, e1600295 (2016).
% HgTexS1-x prediction
\bibitem{Rauch2015} T. Rauch, S. Achilles, J. Henk and I. Mertig, Phys.~Rev.~Lett.~{\bf 114}, 236805 (2015). 
% TaIrTe4
\bibitem{Belopolski2016} I. Belopolski, P. Yu, D.~S. Sanchez, Y. Ishida, T.-R. Chang, S.~S. Zhang, S.-Y. Xu, D. Mou, H. Zheng, 
G. Chang, G. Bian, H.-T. Jeng, T. Kondo, A. Kaminski, H. Lin, Z. Liu, S. Shin and M.~Z. Hasan, arXiv:1610.02013 (2016).

\bibitem{Koepernik2016} K. Koepernik, D. Kasinathan, D.~V. Efremov, S. Khim, S. Borisenko, B. B\"uchner and J. van~den~Brink, Phys.~Rev.~B~{\bf 93}, 201101(R) (2016).
% YbMnBi2
\bibitem{Borisenko2015} S. Borisenko, D. Evtushinsky, Q. Gibson, A. Yaresko, T. Kim, M. N. Ali, B. B\"uchner, M. Hoesch and R.~J. Cava, arXiv:1507.04847 (2015).
% magnetically doped 3D TI layers
\bibitem{Burkov2011} A.~A. Burkov and L. Balents, Phys.~Rev.~Lett.~{\bf 107}, 127205 (2011). 
% magnetically doped Bi2Se3
\bibitem{Cho2011} G.~Y. Cho, arXiv:1110.1939 (2011).
% HgCr2Se4
\bibitem{Xu2011} G. Xu, H. Weng, Z. Wang, X. Dai and Z. Fang, Phys.~Rev.~Lett.~{\bf 107}, 186806 (2011). 
% Hg1-x-yCdxMnyTe
\bibitem{Bulmash2014} D. Bulmash, C.-X. Liu and X.-L. Qi, Phys.~Rev.~B~{\bf 89}, 081106(R) (2014).
% Magnetic Heuslers
\bibitem{Wang2016} Z. Wang, M.G. Vergniory, S. Kushwaha, M. Hirschberger, E.~V. Chulkov, A. Ernst, N.~P. Ong, R.~J. Cava and  B.~A. Bernevig, Phys.~Rev.~Lett.~{\bf 117}, 236401 (2016).
% WSM to Dirac cones
\bibitem{Okugawa2014} R. Okugawa and S. Murakami, Phys.~Rev.~B~{\bf 89}, 235315 (2014).

\bibitem{Grushin2015} A.~G. Grushin, J.~W.~F. Venderbos and J.~H. Bardarson, Phys.~Rev.~B~{\bf 91}, 121109(R) (2015).

\bibitem{Zhang2009} H. Zhang, C.-X. Liu, X.-L. Qi, X.~ Dai, Z. Fang and S.-C. Zhang, Nat.~Phys.~{\bf 5}, 438 (2009).

\bibitem{Liu2010} C.-X. Liu, X.-L. Qi, H. Zhang, X. Dai, Z. Fang and S.-C. Zhang, Phys.~Rev.~B~{\bf 82}, 045122 (2010).

\bibitem{Yang2011} K.-Y. Yang, Y.-M. Lu, and Y. Ran, Phys.~Rev.~B~{\bf 84}, 075129 (2011).

\bibitem{McCormick2016} T.~M. McCormick, I. Kimchi and N. Trivedi, arXiv:1604.03096 (2016).

\bibitem{Xu2015c} Y. Xu, F. Zhang and C. Zhang, Phys.~Rev.~Lett.~{\bf 115}, 265304 (2015). 

\bibitem{Soluyanov2015} A.~A. Soluyanov, D. Gresch, Z. Wang, Q. Wu, M. Troyer, X. Dai and B.~A. Bernevig, Nature~{\bf 527}, 495 (2015).

\bibitem{Sharma2016} G. Sharma, P. Goswami and S. Tewari, arXiv:1608.06625 (2016).

\bibitem{Dwivedi2016} V. Dwivedi and S.~T. Ramamurthy, Phys.~Rev.~B~{\bf 94}, 245143 (2016).
% 2D gapless Thore
\bibitem{Baum2015} Y. Baum, T. Posske, I.~C. Fulga, B. Trauzettel and A. Stern, Phys.~Rev.~Lett.~{\bf 114}, 136801 (2015). 
% time-reversal broken edges
\bibitem{Ma2015} E.~Y. Ma, M.~R. Calvo, J. Wang, B. Lian, M. M\"uhlbauer, C. Br\"une, Y.-T. Cui, K. Lai, W. Kundhikanjana, Y. Yang, M. Baenninger, M. K\"onig, C. Ames, H. Buhmann, P. Leubner, L.~W. Molenkamp, S.-C. Zhang, D. Goldhaber-Gordon, M.~A. Kelly and Z.-X. Shen, Nat.~Comm.~{\bf 6}, 7252 (2015).

\bibitem{Kharitonov2016} M. Kharitonov, S. Juergens and B. Trauzettel, Phys.~Rev.~B~{\bf 94}, 035146 (2016).
% coupling parameters
\bibitem{Michetti2012} P. Michetti, J.~C. Budich, E.~G. Novik and P. Recher, Phys.~Rev.~B~{\bf 85}, 125309 (2012).
% Dirac conductivity
\bibitem{Katsnelson2006} M.~I. Katsnelson, Eur.~Phys.~J.~B~{\bf 51}, 157 (2006).

\bibitem{Tworzydlo2006} J. Tworzyd{\l}o, B. Trauzettel, M. Titov, A. Rycerz, and C.~W.~J. Beenakker, Phys.~Rev.~Lett.~{\bf 96}, 246802 (2006).
% spin polarized arcs
\bibitem{Lv2015c} B.~Q. Lv, S. Muff, T. Qian, Z.~D. Song, S.~M. Nie, N. Xu, P. Richard, C.~E. Matt, N.~C. Plumb, L.~X. Zhao, G.~F. Chen, Z. Fang, X. Dai, J.~H. Dil, J. Mesot, M. Shi, H.~M. Weng and H. Ding, Phys.~Rev.~Lett.~{\bf 115}, 217601 (2015).

\bibitem{Xu2016b} S.-Y. Xu, I. Belopolski, D.~S. Sanchez, M. Neupane, G. Chang, K. Yaji, Z. Yuan, C. Zhang, K. Kuroda, G. Bian, C. Guo, H. Lu, T.-R. Chang, N. Alidoust, H. Zheng, C.-C. Lee, S.-M. Huang, C.-H. Hsu, H.-T. Jeng, A. Bansil, T. Neupert, F. Komori, T. Kondo, S. Shin, H. Lin, S. Jia and M.~Z. Hasan, Phys.~Rev.~Lett.~{\bf 116}, 096801 (2016).

\end{thebibliography}
\end{document}